\documentclass[preprint,noshowkeys,amsmath,amssymb,superscriptaddress,nofootinbib]{revtex4}
\usepackage{graphicx,color,slashed}% Include figure files

\definecolor{green}{cmyk}{1,0,1,0}
\definecolor{pink}{cmyk}{0,0.5,0,0}
 \usepackage{epstopdf}
 \usepackage{slashed}
 \usepackage{ulem}

 \usepackage[bookmarks=true,bookmarksnumbered=true,bookmarkstype=toc, 
colorlinks=true, citecolor=blue, linkcolor=blue]{hyperref} 
\newcommand{\lsim}{\raise0.3ex\hbox{$\;<$\kern-0.75em\raise-1.1ex\hbox{$\sim\;$}}}
\newcommand{\gsim}{\raise0.3ex\hbox{$\;>$\kern-0.75em\raise-1.1ex\hbox{$\sim\;$}}}

\newcommand{\lmlt}{L_\mu^{} - L_\tau^{}}
\newcommand{\mZp}{m_{Z'}}
\newcommand{\Am}{\overline{\mu}}

\usepackage[utf8]{inputenc}

\begin{document}

{\begin{flushright}
{CTP-SCU/2020041} \\
{UME-PP-016}
\end{flushright}}

%%%%%%%%%
\title{Search for $Z'$ pair production from scalar boson decay \\ 
in minimal $U(1)_{L_\mu - L_\tau}$ model  at the LHC}

\author{Takaaki Nomura}
\email{nomura@scu.edu.cn}
\affiliation{College of Physics, Sichuan University, Chengdu 610065, China}

\author{Takashi Shimomura}
\email{shimomura@cc.miyazaki-u.ac.jp}
\affiliation{Faculty of Education, Miyazaki University, Miyazaki, 889-2192, Japan}

\date{\today}

\begin{abstract}
We consider a model with gauged $\lmlt$ symmetry in which the symmetry is spontaneously broken by a scalar field. 
The decay of the scalar boson into two new gauge bosons is studied as a direct consequence of the spontaneous 
symmetry breaking. Then, a possibility of searches for the gauge and scalar bosons through such a decay 
at the LHC experiment is discussed. We consider the case that the mass range of the gauge boson is 
$\mathcal{O}(10)$ GeV, which is motivated by anomalies reported by LHCb. We will show that the signal significance 
of the searches for the gauge boson and scalar boson reach to $3$ and $5$ for the scalar mixing $0.012$ and $0.015$, 
respectively.
\end{abstract}

\maketitle
%{\hypersetup{linkcolor=black}
%\tableofcontents
%}

%%%%%%%%%%%%%%%%%%%%%%%%%%%
\section{Introduction}
%%%%%%%%%%%%%%%%%%%%%%%%%%%
{\it The muon Number minus the tau Number} ($\lmlt$) gauge symmetry has been gaining an attention as an extension of the Standard Model (SM) of particle physics. 
It has been shown that some observed anomalies can be explained by new gauge boson $Z'$ from the symmetry.
One can explain anomalous magnetic moment of muon (muon $g-2$) by the interaction between $Z'$ and muon at one-loop level~\cite{Gninenko:2001hx,Baek:2001kca,Ma:2001md}, 
and  cosmic neutrino spectrum observed at IceCube can be settled by the interaction between $Z'$ and neutrinos~\cite{Aartsen:2014gkd,Araki:2014ona,Kamada:2015era,DiFranzo:2015qea,Araki:2015mya}.
Moreover the anomalies observed in $B \to K^{(*)} \ell^+ \ell^-$ decay ($B$ anomaly) can be explained when we include 
some extra field contents~\cite{Altmannshofer:2014cfa,Altmannshofer:2015mqa,Altmannshofer:2016jzy,Ko:2017yrd,Chen:2017usq, Arcadi:2018tly,Hutauruk:2019crc}. 
Searches for such $Z'$ boson have been studied for on-going and future 
electron collider \cite{Kaneta:2016uyt,Araki:2017wyg,Chen:2017cic,Banerjee:2018mnw,Jho:2019cxq,Iguro:2020rby,Ban:2020uii,Zhang:2020fiu}, 
hadron collider \cite{Harigaya:2013twa}, 
beam dump \cite{Gninenko:2018tlp,Gninenko:2020hbd}, 
neutrino beam experiments \cite{Altmannshofer:2019zhy,Ballett:2019xoj,Shimomura:2020tmg}
and meson decay \cite{Ibe:2016dir}, atmospheric neutrino observations \cite{Ge:2017poy}.

Although the favored range of the $Z'$ mass is different for explaining each anomaly, the gauge boson is required to be massive to explain them.
Therefore the origin of the gauge boson mass should be implemented in full models.
Spontaneous symmetry breaking is one of the natural mechanisms to generate the gauge boson mass.  
Then, the scalar sector of the SM also needs to be enlarged by introducing new scalar field(s). 
Observations of such the scalar boson as well as the gauge boson will provide important hints of the origin of particle 
mass and interaction.

In spontaneous symmetry breaking mechanism,  the gauge boson $Z'$ acquires a mass after a new scalar field breaks the $\lmlt$ gauge symmetry by developing a vacuum expectation value (vev). 
The gauge boson mass term emerges from the interaction term of two scalar bosons ($\phi$) and two gauge bosons 
($\phi-\phi-Z'-Z'$) by replacing the scalar field with its vev. 
Then, as a direct consequence of the spontaneous symmetry breaking, the interaction of the massive scalar boson with 
two gauge bosons ($\phi-Z'-Z'$) also emerges. 
By this interaction term, the scalar boson can decay into two gauge bosons if kinematically allowed. 
Such a decay is a clear signature of the spontaneous symmetry breaking or the mass generation.
An interesting feature of this interaction is that the decay of $\phi$ 
into two $Z'$ can be enhanced when the gauge boson is lighter than the scalar boson. 
This is because the longitudinal component of $Z'$, which is the Nambu-Goldstone mode of $\phi$, 
is inversely proportional to its mass. 
In the case of very light $Z'$ compared with $\phi$, 
the decay width into two $Z'$ can dominate other decay widths. Then, $\phi$ dominantly decays into two $Z'$.

The scalar boson, on the other hand, can be produced at collider experiments through mixing with the SM Higgs boson. 
In general such a mixing is expected to exist because  quartic terms of the SM Higgs boson and $\phi$ are allowed by the symmetry. The mixing 
can be simultaneously generated through the quartic terms after the spontaneous symmetry breaking. 
Once the scalar boson is produced, it can dominantly decay into the lighter gauge bosons when the decay widths into the SM particles are suppressed.
In the case that the gauge boson further decays into charged SM particles, the mass of the scalar boson and gauge boson can be reconstructed from the energies and momenta of the SM particles. 

Searches for the scalar boson and gauge boson through the above decay at the LHC and ILC have been studied for the $\lmlt$ gauge boson with mass lighter than 200 MeV in ref.~\cite{Nomura:2018yej} 
where the mass region is favored to explain muon $g-2$. 
We found that such a light gauge boson decaying into charged leptons is difficult to identify in the LHC experiments 
while the scalar and gauge boson masses can be reconstructed at the ILC experiment using energy-momentum conservation.
Thus, the confirmation of the $\phi-Z'-Z'$ interaction will be possible at the ILC. 
Similar studies for dark photon from a light scalar boson decay at the FASER experiment have been done in \cite{Araki:2020wkq}.
Since the above studies are focused on light gauge bosons, it is also important to search for the signal when 
the $\lmlt$ gauge boson is heavier than $\mathcal{O}(1)$ GeV, which is motivated to explain $B$ anomaly.
In this paper, we consider such heavier $Z'$ with mass $\mathcal{O}(10)$ GeV 
in a minimal $\lmlt$ model, and study a possibility of search for both the gauge and scalar bosons at the LHC.

This paper is organized as follows.
In section II, we review a minimal gauged $L_\mu - L_\tau$ model summarizing interactions and mass spectrum.
In section III, we show decay widths of new bosons and the SM-like Higgs bosons.
In section IV, we discuss experimental constraints for new gauge coupling, $Z'$ mass and scalar mixing. 
In section V, we carry out numerical simulation for our signal and background, and the estimate discovery significance.
Our conclusion is given in section VI.

%%%%%%%%%%%%%%%%%%%%%%%%%%%%%%
\section{Minimal $L_\mu - L_\tau$ Model} \label{sec:model}
%%%%%%%%%%%%%%%%%%%%%%%%%%%%%%
We start our discussion with introducing a minimal gauged $\lmlt$ model, where $L_\mu$ and $L_\tau$ represent 
the muon $(\mu)$ and tau $(\tau)$ number, respectively. 
The gauge symmetry of the model is defined by adding $U(1)_{L_\mu - L_\tau}$ to that of the SM.  
Under the $\lmlt$ symmetry, only muon and tau flavour leptons among the SM particles are charged. 
As a minimal setup, the scalar sector is also extended by introducing one complex scalar, $\varphi$, which is 
charged under the $\lmlt$ symmetry and is singlet under the SM gauge symmetry\footnote{When we introduce right-handed neutrinos, neutrino masses and 
mixing can be generated via Yukawa coupling with $\varphi$ . 
However, in such a case, neutrino mass spectrum has a strong tension with Planck observation \cite{Aghanim:2018eyx,Asai:2019ciz}. The constraint can be evaded by introducing other scalar bosons. In that case, our results will be applicable to the scalar boson which mainly 
consists of the SM singlet one.  From this reason, 
we do not consider neutrino masses and mixing, and focus on the singlet scalar $\varphi$ in this work.}.
The gauge charge assignment for the fermions and scalars under the weak $SU(2)_W$ and 
the hypercharge $U(1)_Y$ symmetries is shown in Table \ref{tab:charge}. 
In the table, $Q_L$ and $u_R,~d_R$ are $SU(2)_W$ doublet left-handed  quarks and singlet right-handed 
up-type, down-type quarks, and $l_\alpha$ and $\alpha_R$ $(\alpha = e,~\mu,~\tau)$ are $SU(2)_W$ doublet left-handed 
and  singlet right-handed charged leptons, respectively. Here, $e$ stands for electron flavour.
The $SU(2)_W$ doublet scalar, $H$, is responsible for the EW symmetry breaking while the singlet  scalar, $\varphi$, 
is for the $\lmlt$ symmetry breaking. The strong interaction part of the model is the same as that of the SM.
\begin{table}[t]
  \begin{center}
    \begin{tabular}{|c|c|c|c|c|c|c|c|c|c||c|c|}\hline
      &
      $~~Q_L~~$ & 
      $~~u_R~~$ & 
      $~~d_R~~$ & 
      $~~l_e~~$ & 
      $~~l_\mu~~$ & 
      $~~l_\tau~~$ & 
      $~~e_R~~$ & $~~\mu_R~~$ & $~~\tau_R~~$ &
      $~~H~~$ & $~~\varphi~~$ \\ \hline
      $~~SU(2)_W~~$ & $\bf{2}$ & $\bf{1}$ & $\bf{1}$ & 
      $\bf{2}$ & $\bf{2}$ & $\bf{2}$ & $\bf{1}$ &$\bf{1}$ & $\bf{1}$ & $\bf{2}$ & $\bf{1}$ \\ \hline
      $~~U(1)_Y~~$ & $\frac{1}{6}$ & $\frac{2}{3}$ & $-\frac{1}{3}$ & 
      				$-\frac{1}{2}$ & $-\frac{1}{2}$ & $-\frac{1}{2}$ & $-1$ &$-1$ & $-1$ & $\frac{1}{2}$ & $0$ \\ \hline
      $~~U(1)_{L_\mu -L_\tau}~~$ & $0$ & $0$ & $0$ & $0$ & $1$ & $-1$ & $0$ &$1$ & $-1$ & $0$ & $1$ \\ \hline
    \end{tabular}
  \end{center}
  \caption{The gauge charge assignment of a gauged minimal $U(1)_{L_\mu - L_\tau}$ model. }
  \label{tab:charge}
\end{table}

%%%%%%%%%%%%%%%%%%%%%%%%%%%
\subsection{Lagrangian} \label{subsec:lagrangian}
%%%%%%%%%%%%%%%%%%%%%%%%%%%

The Lagrangian of the model takes the form of 
\begin{align}
\mathcal{L} =& \mathcal{L}_{\mathrm{SM}}- \frac{1}{4} Z'_{\mu \nu} Z'^{\mu \nu} 
+ g' Z'_\mu J^\mu_{\lmlt} - \frac{\epsilon}{2} B_{\mu \nu} Z'^{\mu \nu}  
+|D_\mu \varphi|^2 - V, \label{eq:Lagrangian} 
\end{align}
where $\mathcal{L}_{\mathrm{SM}}$ represents the SM Lagrangian except for the Higgs potential, and 
$Z'$ and $B$ stand for the $\lmlt$ and hypercharge gauge  boson and its field strength in interaction basis, 
respectively. The $\lmlt$ gauge coupling constant is denoted as $g'$ and its gauge current $J_{\lmlt}^\mu$ is 
given by
\begin{align}
J^\mu_{\lmlt} = \bar l_\mu \gamma^\mu l_\mu + \bar \mu_R \gamma^\mu \mu_R - \bar l_\tau \gamma^\mu l_\tau - \bar \tau_R \gamma^\mu \tau_R, \label{eq:current} 
\end{align}  
The gauge kinetic mixing term between two $U(1)$ gauge bosons is allowed by the symmetry, which is parametrized by 
$\epsilon$. 
The covariant derivative for the kinetic term of $\varphi$ is given by 
\begin{align}
D_\mu = \partial_\mu - i g' Z'_\mu.
\end{align}
The scalar potential $V$ is given by 
\begin{align}
V =   -\mu_H^2 H^\dagger H - \mu_\varphi^2 \varphi^* \varphi + \frac{\lambda_H}{2} (H^\dagger H)^2 + \frac{\lambda_\varphi}{2} (\varphi^* \varphi)^2 + \lambda_{H \varphi} (H^\dagger H)(\varphi^* \varphi). \label{eq:scalar-potential}
\end{align} 
where $\mu_H$ and $\mu_\varphi$ are the mass parameters of $H$ and $\varphi$, and $\lambda_H,~\lambda_\varphi$ and $\lambda_{H\varphi}$  are the quartic couplings, 
respectively.

In the following discussion, we assume that the values of $\mu_H^2$ and $\mu_\varphi^2$ as well as $\lambda_H,~\lambda_\varphi$ and $\lambda_{H\varphi}$, are 
set appropriately so that the scalar fields develop vacuum expectation values (vevs) to break the EW and $\lmlt$ symmetries on a stable vacuum.
Furthermore, to simplify our analysis, we assume that the gauge kinetic mixing term is absent at tree-level.
Even though, such a kinetic mixing can be generated radiatively via muon and tau loops. 
The loop-induced kinetic mixing with photon is given by
\begin{align}
\epsilon(q^2) &= \frac{8 e g'}{(4 \pi)^2} \int_0^1 dx x(1-x) \log\left[
\frac{m_\tau^2 -x (1-x) q^2}{m_\mu^2 -x (1-x) q^2}
\right], \label{eq:loop-induced-mix}
\end{align}
where $q^2$ is the four momentum squared carried by $Z'$ and $e$ is the electric charge of proton.\footnote{We use the same symbol $e$ for electron flavour and electric charge.} 
Masses of muon and tau leptons are denoted as $m_{\mu}$ and $m_{\tau}$, respectively. 
When $Z'$ is on-shell, $q^2$ can be replaced with the $Z'$ boson mass, $m_{Z'}^2$. 
%
%%%%%%%%%%%%%%%%%%%%%%%%%
\begin{figure}[t]
\begin{center}
 \includegraphics[height=55mm]{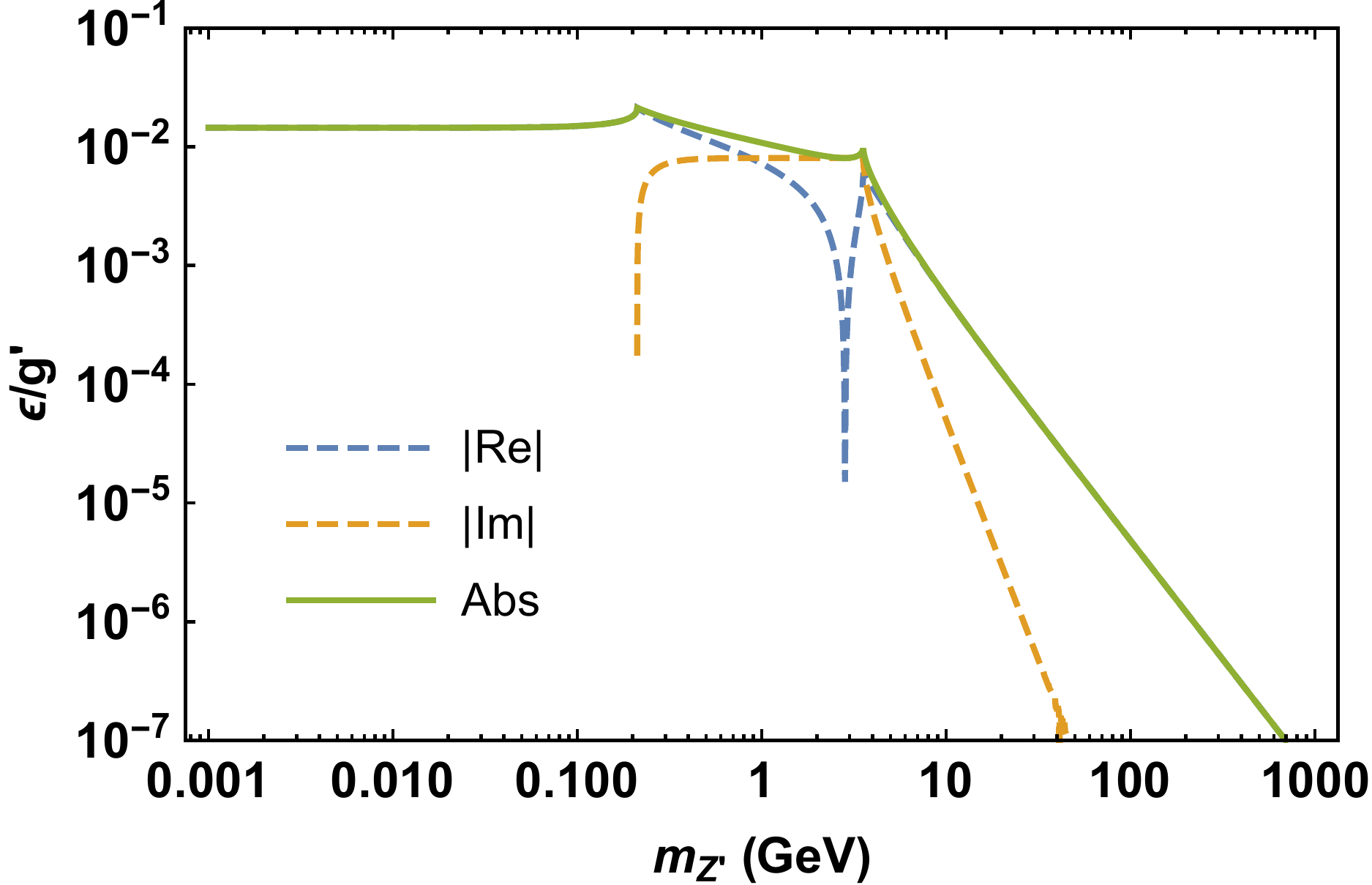}
\end{center}
\caption{ 
Loop induced kinetic mixing normalized by $g'$ in $L_\mu - L_\tau$ model. 
The blue and orange dashed curves represent the absolute value of the real and imaginary part of 
$\epsilon(q^2 = \mZp^2)$, and the green solid one is the absolute value of $\epsilon(q^2 = \mZp^2)$.
}
\label{fig:loop-induced-kinetic-mixing}
\end{figure}
%%%%%%%%%%%%%%%%%%%%%%%%%
%
Figure \ref{fig:loop-induced-kinetic-mixing} shows the ratio, $\epsilon(\mZp^2)/g'$, as a function of $\mZp$.  
Blue and orange curves represent the absolute value of the real part and imaginary part, and green one represents the absolute value 
of the ratio, respectively. The imaginary part is comparable to or larger than the real part for 
$2 m_\mu \leq \mZp \leq 2 m_\tau$. Above a tau pair production threshold, both of the real and imaginary parts monotonically 
decrease as $\mZp$ increases. The imaginary part decreases faster than the real part does. 
This is because the imaginary part is inversely proportional to $q^4$ while the real part is to $q^2$.
The mass range of our interest is $20$ GeV $\lsim \mZp \lsim 60$ GeV, and $\epsilon(\mZp^2)/g'$ varies from $10^{-4}$ 
to $10^{-5}$ in this range. Therefore the loop-induced kinetic mixing is negligible compared with $g'$ 
in the $Z'$ decays into $\mu \bar{\mu}$ and $\tau \bar{\tau}$ pairs. 
On the other hand, $Z'$ can decay into electron-positron ($e^- e^+)$ and quark-antiquark ($q\bar{q})$ pairs only through this mixing. 
However, as we will show later, the branching ratios of $Z' \to \mu \bar{\mu}/\tau \bar{\tau}$ are 
$\mathcal{O}(10^4 - 10^{10})$ times larger than that of $Z' \to e^- e^+$ and $q\bar{q}$.
The expected numbers of $Z'$ decay into these pairs are negligible.
Therefore, we can safely ignore the gauge kinetic mixing in our analysis of the scalar production and $Z'$ decays.
%

%%%%%%%%%%%%%%%%%%%%%%%%%%%%%%%%%%%%%%%%%%%%%%%
\subsection{Gauge Boson Mass} \label{subsec:gauge-sector}
%%%%%%%%%%%%%%%%%%%%%%%%%%%%%%%%%%%%%%%%%%%%%%%
The gauge boson acquire masses after the EW and $\lmlt$ symmetries are spontaneously broken by the vevs 
of the scalar fields, $H$ and $\varphi$, 
\begin{align}
\langle H \rangle = 
\frac{v}{\sqrt{2}}, \quad
\langle \varphi \rangle = \frac{v_\varphi}{\sqrt{2}}.
\label{eq:vevs}
\end{align}
Inserting Eq.~\eqref{eq:vevs} into the kinetic term of $H$ and $\varphi$, the mass terms of the gauge bosons are 
obtained as
\begin{align}
\mathcal{L}_{\mathrm{gauge, mass}} =\frac{1}{2}m_{Z}^2 Z_\mu Z^\mu + m_{W}^2 W^+_\mu W^{-\mu} 
+ \frac{1}{2} m_{Z'}^2 Z'_\mu Z'^\mu,
\label{eq:gauge-mass-term}
\end{align}
where $A,~Z$ and $W^\pm$ are the SM photon, $Z$ and $W$ boson are defined by
\begin{subequations}
\begin{align}
A_\mu &= \sin\theta_W W^3_\mu + \cos\theta_W B_\mu, \\
Z_\mu &= \cos\theta_W W^3_\mu - \sin\theta_W B_\mu, \\
W^\pm_\mu &= \frac{1}{\sqrt{2}} (W^1_\mu \mp i W^2_\mu), 
\end{align}
\label{eq:gauge-SM}
\end{subequations}
and the masses are given by 
\begin{subequations}
\begin{align}
m_{Z}^2 &= \frac{1}{2} (g_1^2 + g_2^2) v^2, \\
m_W^2 & = g_2^2 v^2, \\
m_{Z'}^2 &= g'^2 v_\varphi^2.
\end{align}
\label{eq:gauge-masses}
\end{subequations}
Here $W^a_\mu~(a=1,2,3)$ and $g_2$ are the $SU(2)_W$ gauge bosons and coupling constant, respectively, 
and $\theta_W$ is the Weinberg angle defined by $\sin\theta_W =g_1/\sqrt{g_1^2+g_2^2}$ in the SM. 
It should be noticed that $A,~Z$ and $Z'$ do not mix because we ignore the gauge kinetic mixing.

%%%%%%%%%%%%%%%%%%%%%%%%%%%%%%%%%%%%%%%%%%%%%%%
\subsection{Scalar Boson Mass and Mixing} \label{subsec:scalar-sector}
%%%%%%%%%%%%%%%%%%%%%%%%%%%%%%%%%%%%%%%%%%%%%%%
The scalar bosons also acquire their masses after spontaneous symmetry breaking. 
We expand $H$ and $\varphi$ around its vev as 
\begin{align}
H = 
\begin{pmatrix}
w^+ \\
\frac{1}{\sqrt{2}} (v + \tilde{h} + i \xi ) 
\end{pmatrix}, \quad
\varphi &= \frac{1}{\sqrt{2}} (v_\varphi + \tilde{\phi} + i \eta),
\label{eq:scalar-fields}
\end{align}
where $\tilde{h}$ and $\tilde{\phi}$ are CP-even scalar bosons as physical degree of freedom while $w^+$, 
$\xi$ and $\eta$ are the Nambu-Goldstone bosons which are absorbed by the weak gauge boson $Z,~W$ and $Z'$.
The scalar boson masses are obtained by inserting Eq.~\eqref{eq:scalar-fields} into the potential.
The mass matrix of the CP-even scalar bosons is given by
\begin{align}
M^2_{\mathrm{even}} = 
\begin{pmatrix}
\lambda_H v^2 & \lambda_{H\varphi} v v_\varphi \\
\lambda_{H\varphi} v v_\varphi & \lambda_\varphi v_\varphi^2
\end{pmatrix},
\end{align}
where we used the stationary conditions,
\begin{align}
\frac{\partial V}{\partial H}= \frac{\partial V}{\partial \varphi} =0.
\end{align}
The mass matrix can be diagonalized by an orthogonal matrix $U_{\mathrm{even}}$ as
\begin{align}
 U^T_{\mathrm{even}} M^2_{\mathrm{even}} U_{\mathrm{even}} = \mathrm{diag}(m_h^2, m_H^2),
\end{align}
where 
\begin{align}
U_{\mathrm{even}} = 
\begin{pmatrix}
\cos\alpha & \sin\alpha \\
-\sin\alpha & \cos\alpha
\end{pmatrix}.
\end{align}
The mass eigenvalues are obtained as
\begin{subequations}
\begin{align}
m_h^2 &= \lambda_H v^2 c_\alpha^2 + \lambda_\varphi v_\varphi^2 s_\alpha^2 
	+ 2 \lambda_{H\varphi} v v_\varphi s_\alpha c_\alpha, \\
m_\phi^2 &= \lambda_\varphi v_\varphi^2 s_\alpha^2 + \lambda_H v^2 s_\alpha^2 
	- 2 \lambda_{H\varphi} v v_\varphi s_\alpha c_\alpha, 
\end{align}
\label{eq:scalar-masses}
\end{subequations}
where the scalar mixing angle $\alpha$ is expressed as
\begin{align}
\tan 2\alpha = \frac{2 \lambda_{H \varphi} v v_\varphi}{\lambda_H v^2 - \lambda_\varphi v_\varphi^2}. \label{eq:scalar-mixing}
\end{align}
The corresponding mass eigenstates are given by
\begin{align}
\begin{pmatrix}
h \\
\phi
\end{pmatrix}
=
U^T_{\mathrm{even}}
\begin{pmatrix}
\tilde{h} \\
\tilde{\phi}
\end{pmatrix}
=
\begin{pmatrix}
\cos\alpha & -\sin\alpha \\
\sin\alpha & \cos\alpha
\end{pmatrix}
\begin{pmatrix}
\tilde{h} \\
\tilde{\phi}
\end{pmatrix}.
\label{eq:scalar-eigenstates}
\end{align}
Note that $h$ becomes the SM Higgs boson in the limit of $\alpha \to 0$.

In this work, we employ the masses and mixing angle as input parameters and express the quartic coupling in terms 
of the inputs. From Eqs.~\eqref{eq:scalar-masses} and \eqref{eq:scalar-mixing}, the quartic couplings can be 
given by
\begin{subequations}
\begin{align}
\lambda_H &= \frac{1}{2 v^2}\big( m_h^2 + m_\phi^2 + (m_h^2 - m_\phi^2) \cos 2\alpha \big), \\
\lambda_\varphi &= \frac{1}{2 v_\varphi^2}\big( m_h^2 + m_\phi^2 - (m_h^2 - m_\phi^2) \cos 2\alpha \big), \\
\lambda_{H\varphi} &= \frac{1}{2 v v_\varphi} (m_h^2 - m_\phi^2) \sin 2\alpha.
\end{align} 
\label{eq:quartic_coup}
\end{subequations}
where $v$ and $v_\varphi$ are given by $m_W/g_2$ and $m_{Z'}/g'$ from Eqs.~\eqref{eq:gauge-masses}, respectively. 
Note that $\lambda_H$ and $\lambda_\varphi$ are always positive because positive mass squared of $h$ and $\phi$ 
while $\lambda_{H\varphi}$ can be nagative.

%%%%%%%%%%%%%%%%%%%%%%%%%%%%%%%%%%%%%%%%%%%%%%%
\section{Decay Widths} \label{sec:decay-width}
%%%%%%%%%%%%%%%%%%%%%%%%%%%%%%%%%%%%%%%%%%%%%%%
In this seciton, we present the partial decay widths of $\phi$ and $Z'$ for estimating the signal events 
while those of the SM-like Higgs, $h$, for constraints. 

%%%%%%%%%%%%%%%%%%%%%%%%%%%%%%%%%%%%%%%%%%%%%%%
\subsection{Decays of $\phi$} \label{subsec:decay-phi}
%%%%%%%%%%%%%%%%%%%%%%%%%%%%%%%%%%%%%%%%%%%%%%%
In our analysis the signal of new bosons is obtained from the decays $\phi \to Z'Z'$ followed by $Z' \to \mu \mu$. 
Note also that $\phi$ can decay into the SM fermions $(f)$, the SM-like Higgs $(h)$ 
and the SM vector bosons $(V)$ depending on its mass,  
%%%
\begin{subequations}
\begin{align}
\phi &\to f \overline{f}, \label{eq:phi-ff} \\
\phi &\to h h, \label{eq:phi-hh}\\
\phi &\to V V, \label{eq:phi-VV}\\
\phi &\to V V^\ast \to V f \overline{f} (f \overline{f'}). \label{eq:phi-VVast}
\end{align}
\end{subequations}
The partial widths of the decays Eq.~\eqref{eq:phi-VVast} can be sizable when $m_\phi$ is smaller than $2m_W$ and/or $2m_Z$, 
although these are next-leading order.

The relevant interaction Lagrangian for the decays is given by
\begin{align}
\mathcal{L}_{\phi-\mathrm{int}} &= - \frac{\sin\alpha}{v} m_f \overline{f} f \phi -\lambda_{\phi 2h} h^2 \phi \nonumber \\
	&\quad +\left(
	-2 \frac{m_W^2}{v}\sin\alpha W^+_\mu W^{-\mu} - \frac{m_Z^2}{v} \sin\alpha Z_\mu Z^\mu 
	+ \frac{\mZp^2}{v_\varphi} \cos\alpha Z'_\mu Z'^\mu 
	\right) \phi,
\label{eq:int-lag-phi}
\end{align}
where $m_f$ is the mass of fermion, $f$, and 
\begin{align}
\lambda_{\phi 2h} &= \frac{\sin 2\alpha (2 m_h^2 + m_\phi^2) ( \cos \alpha v_\varphi - \sin \alpha v)}{4 v v_\varphi}.
\end{align}
The interaction of $\phi$ with $Z$ and $Z'$ is absent without the tree-level gauge kinetic mixing. 
It can be generated at loop-level, but it is much suppressed as we have explained. Therefore, we do not consider 
the decay $\phi \to Z Z'$. 
Then, using Eq.~\eqref{eq:int-lag-phi}, the partial decay width of $Z'Z'$ mode is obtained as 
\begin{align}
\Gamma(\phi \to Z'Z') &= \frac{\mZp^4 \cos^2\alpha}{8 \pi v_\varphi^2 m_\phi}\beta(x_{Z'}) 
	\left[ 2 + \frac{m_\phi^4}{4 \mZp^4} \left( 1 - \frac{2 \mZp^2}{m_\phi^2} \right)^2 \right], \label{eq:phi-decay-zpzp}
\end{align}
and those into the SM particles are given by
\begin{subequations}
\begin{align}
 \Gamma(\phi \to f \overline{f}) &= \frac{N_c \sin^2 \alpha}{8 \pi} \frac{m_f^2}{v^2} \beta(x_f)^3 m_\phi, \\
\Gamma(\phi \to hh) &= \frac{\lambda_{2h \phi}^2}{8 \pi m_\phi} \beta(x_h), \\
\Gamma(\phi \to WW) &= \frac{m_W^4 \sin^2\alpha}{4 \pi v^2 m_\phi}\beta(x_W) 
	\left[ 2 + \frac{m_\phi^4}{4 m_W^4} \left( 1 - \frac{2 m_W^2}{m_\phi^2} \right)^2 \right],  \\
\Gamma(\phi \to ZZ) &= \frac{m_Z^4 \sin^2\alpha}{8 \pi v^2 m_\phi}\beta(x_Z) 
	\left[ 2 + \frac{m_\phi^4}{4 m_Z^4} \left( 1 - \frac{2 m_Z^2}{m_\phi^2} \right)^2 \right], 
\end{align}
\end{subequations}
where $N_c = 3~(1)$ is the color factor for quarks (leptons) and $\beta(x) = \sqrt{1 - 4 x}$ with 
$x_i = m_i^2/m_\phi^2$~$(i=Z',f,h,W,Z)$. 
For the decay modes in Eq.~\eqref{eq:phi-VVast}, the decay widths are given by summing over fermions with neglecting their masses,
\begin{subequations}
\begin{align}
\Gamma(\phi \to W W^\ast \to W f \bar{f'}) &= \frac{3 m_W^4 \sin^2\alpha}{2^{6} \pi^3 v^4} m_\phi S(x_W), \\
\Gamma(\phi \to Z Z^\ast \to W f \bar{f'}) &= \frac{3 m_Z^2 m_W^2\sin^2 \alpha}{2^6 \pi^3 v^4 \cos^2\theta_W} m_\phi R_W S(x_Z) 
\end{align}
\end{subequations}
where $\theta_W$ is the Weinberg angle, and 
\begin{subequations}
\begin{align}
S(x) &= 47 x^2 - 60 x + 15 - \frac{2}{x} - 3 (4 x^2 - 6 x + 1) \log x \nonumber \\
 &\quad +  \frac{6( 20 x^2 - 8 x + 1) }{\sqrt{4x-1}} \cos^{-1} \left( \frac{3x-1}{2 x^{3/2} } \right), \\
 R_W &= \frac{7}{12} - \frac{10}{9} \sin^2\theta_W + \frac{40}{27} \sin^4\theta_W.
\end{align}
\end{subequations}

%%%%%%%%%%%%%%%%%%%%%%%%%%%%%%%%%%%%%%%%%%%%%%%%
\begin{figure}[t]
\centering
	\begin{minipage}{1.0\textwidth}
		\centering
		\begin{minipage}{0.49\textwidth}
			\includegraphics[width=0.9\textwidth]{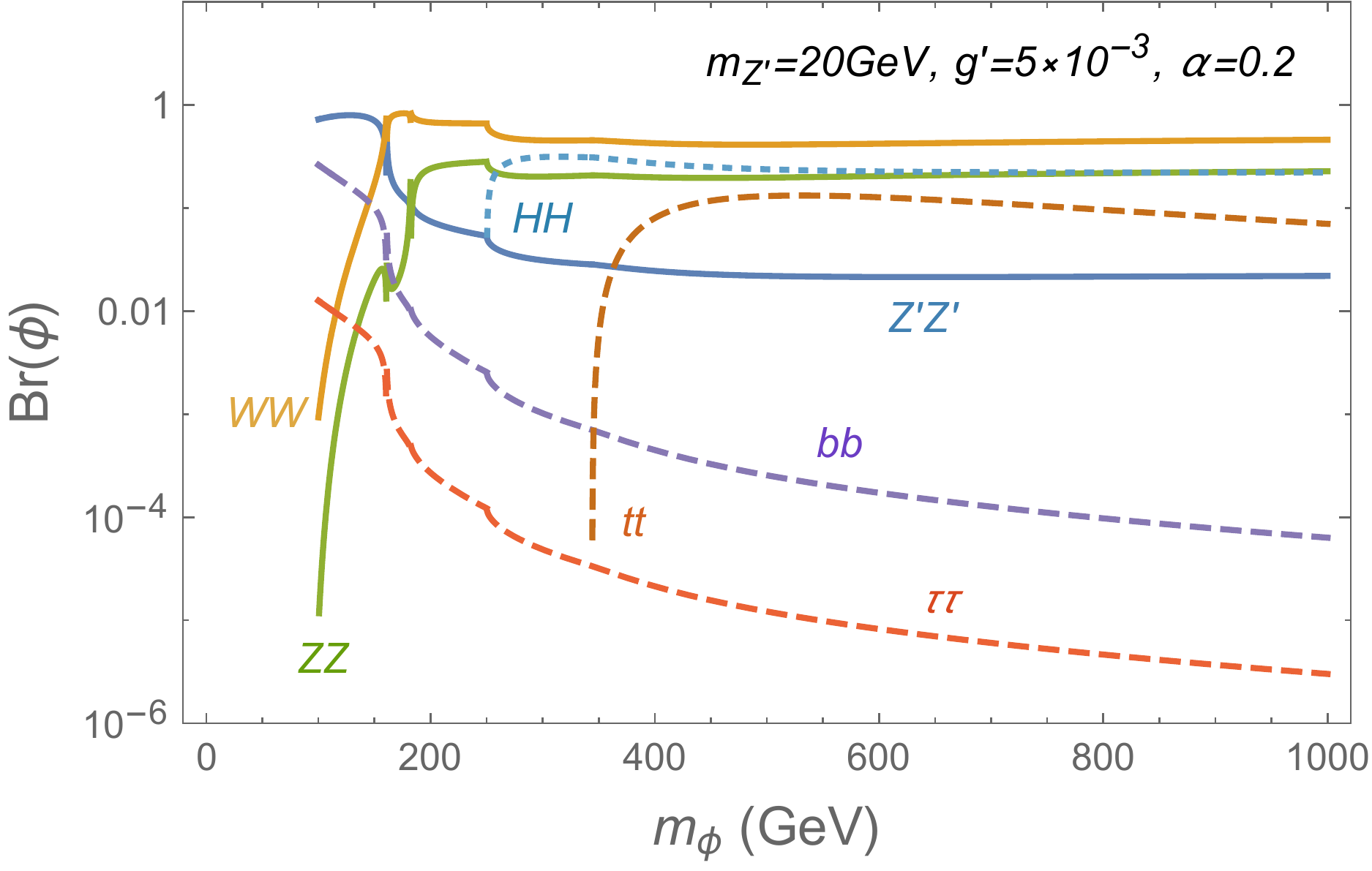} 
		\end{minipage}
		\centering
		\begin{minipage}{0.49\textwidth}
			\includegraphics[width=0.9\textwidth]{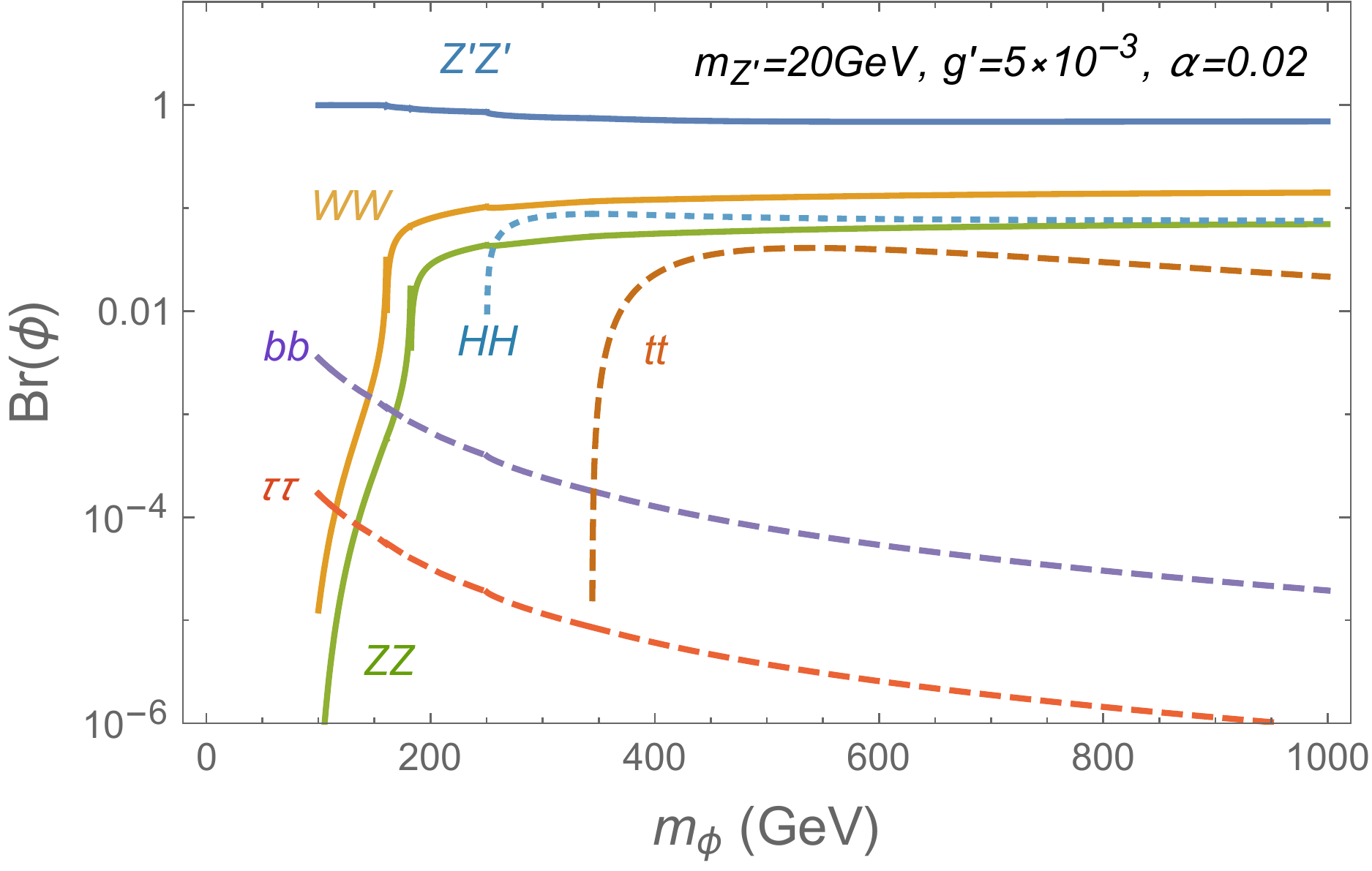} 
		\end{minipage}
	\end{minipage}
\vspace{5mm} \mbox{}
	\begin{minipage}{1.0\textwidth}
		\centering
		\begin{minipage}{0.49\textwidth}
			\includegraphics[width=0.9\textwidth]{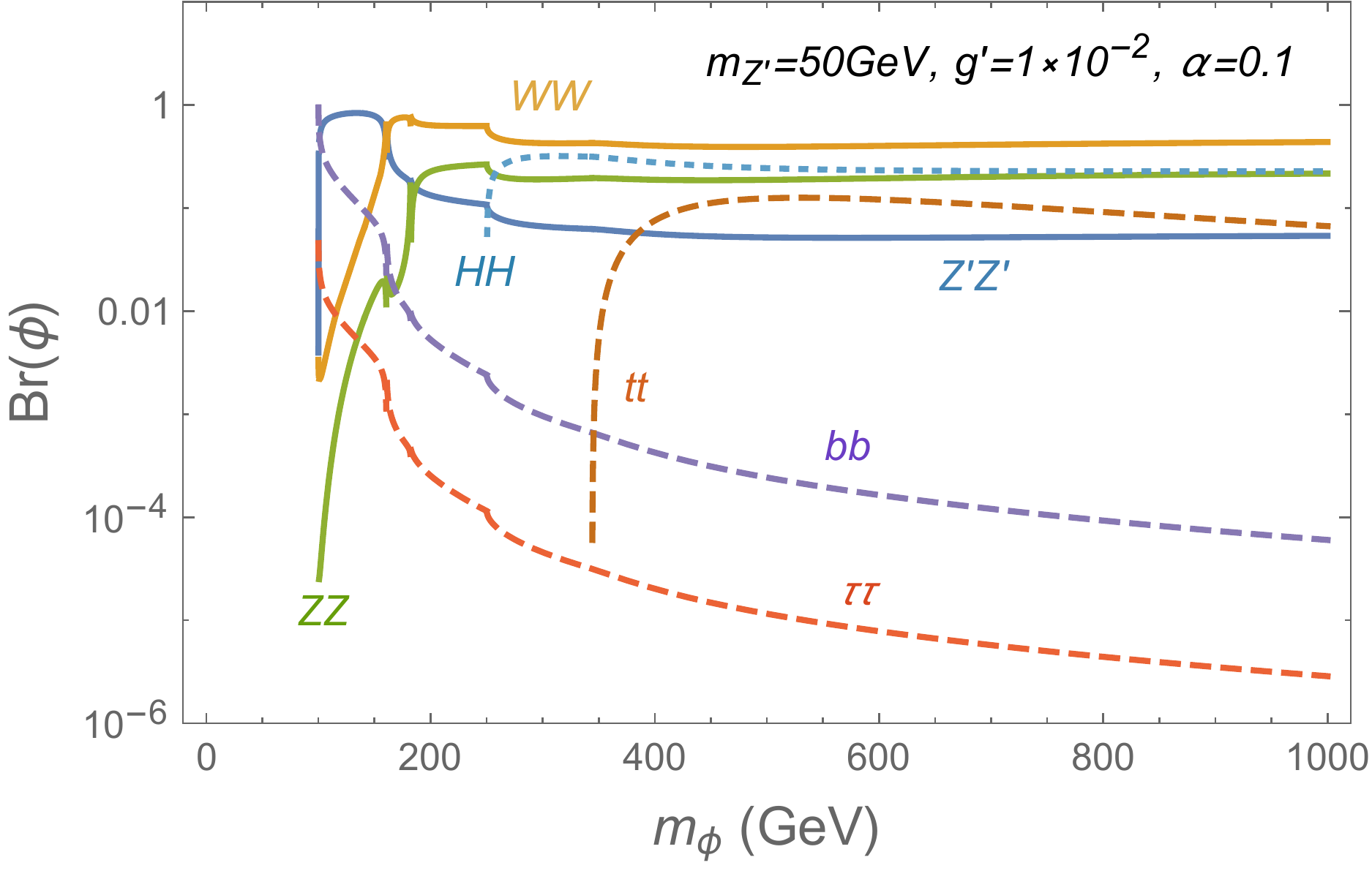} 
		\end{minipage}
		\centering
		\begin{minipage}{0.49\textwidth}
			\includegraphics[width=0.9\textwidth]{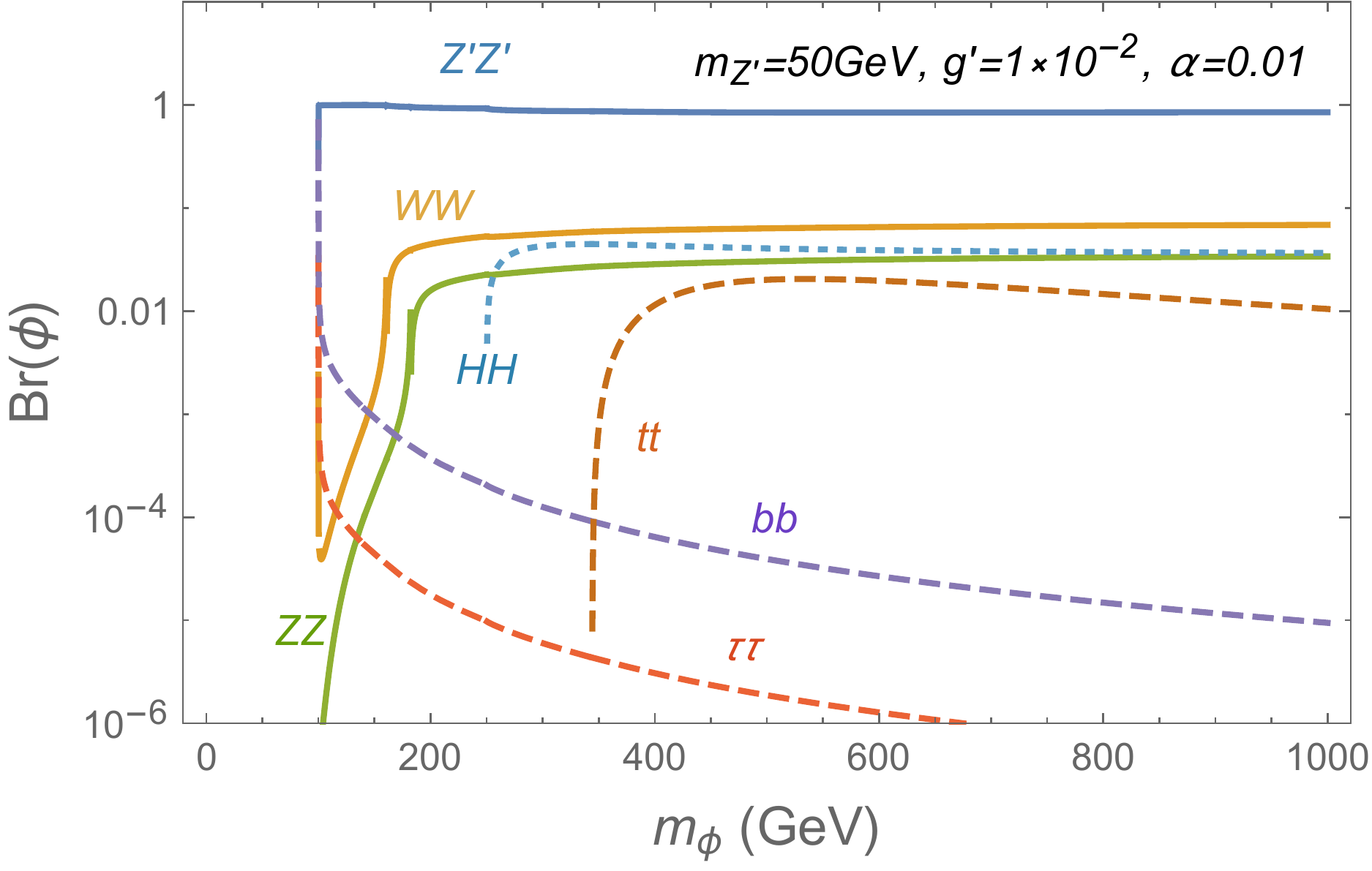} 
		\end{minipage}
	\end{minipage}
\caption{The branching ratios of $\phi$ decays as a function of $m_\phi$. The final states are indicated near each curves. }
\label{fig:br-phi}
\end{figure}
%%%%%%%%%%%%%%%%%%%%%%%%%%%%%%%%%%%%%%%%%%%%%%%%%%%
Figures \ref{fig:br-phi} shows numerical results of the decay branching ratios of $\phi$, $\mathrm{Br}(\phi)$, 
as a function of $m_\phi$. The $Z'$ boson mass is taken to be $20$ and $40$ GeV in top and bottom panels 
as reference values, respectively.
Other parameters, $g'$ and $\alpha$, for each $\mZp$ are indicated in each panels. These parameters are chosen so as to avoid the present experimental bounds on $\sin\alpha \lsim 0.25$ and/or $h \to Z'Z' \to 4l$, which we will discuss in the next section. In the figures, blue, green and yellow solid curves represent the branching ratios 
into $Z'Z',~ZZ$ and $WW$, and light blue, brown, purple and red dashed curves represent those into $hh$, top $(tt)$, 
bottom $(bb)$, and tau pairs, respectively.
From the figures, the branching ratio of $\phi \to Z'Z'$ is larger for smaller value of $\alpha$ to each $\mZp$. 
This is simply because the interaction of $\phi$ with $Z'$ comes from the $\lmlt$ gauge interaction and 
hence is proportional to $\cos\alpha$, as presented in Eq.~\eqref{eq:int-lag-phi}. 
On the other hand, the interactions with the SM particles, $ZZ,~WW$ and $hh,~tt,~bb,~\tau\tau$ is proportional to $\sin\alpha$ because it is generated through the mixing with the SM Higgs. 
Thus, the decay widths into $Z'Z'$ can be dominat for smaller $\alpha$.
However, in such a case, the production cross section of $\phi$ becomes much suppressed since $\phi$ is 
mainly produced via gluon-fusion which is also proportional to $\sin^2 \alpha$. 
Therefore, there are certain parameter regions where $\phi$ and $Z'$ will be detected at the LHC when both production cross section and BR of $Z'Z'$ mode is sizable.
We will search for such parameter regions performing numerical simulation analysis in Sec.~\ref{sec:search}.

%%%%%%%%%%%%%%%%%%%%%%%%%%%%%%%%%%%%%%%%%%%%%%%
\subsection{Decays of $Z'$} \label{subsec:decay-Zprime}
%%%%%%%%%%%%%%%%%%%%%%%%%%%%%%%%%%%%%%%%%%%%%%%
In the mass range of our interest, $\mZp < 60$ GeV, the $Z'$ boson decays mainly into muon and tau lepton pairs 
through the $\lmlt$ gauge interaction while it decays partly into electron and quarks 
through the loop-induced kinetic mixing.  
Although these decay widths are suppressed, we include the effects of the loop-induced kinetic 
mixing for completeness only in this subsection. 

The relevant interaction Lagrangian with the $Z'$ decays into fermion $f$ is given by
\begin{align}
\mathcal{L}_{Z'-\mathrm{int}} =\frac{1}{2} \bar{f} \gamma^\mu (v'_f - a'_f \gamma_5) f Z'_\mu,
\end{align}
where 
\begin{subequations}
\begin{align}
v'_f &= 2 e Q_f U_{13} + \frac{g_2}{\cos\theta_W}(T_{3f} - 2 Q_f \sin^2\theta_W) U_{23} + 2 g' X_f U_{33}, \\
a'_f &= \frac{g_2}{\cos\theta_W}(T_{3f} ) U_{23}.
\end{align}
\label{eq:coup-Zprime-ff}
\end{subequations}
In Eqs.~\eqref{eq:coup-Zprime-ff}, $Q_f,~T_{3f}$ and $X_f$ are the electric, weak and $\lmlt$ charges 
of $f$, respectively. The elements of the gauge mixing matrix, $U$, is given by
\begin{align}
U_{13} = \epsilon r \cos\theta_W \cos\chi, ~~U_{23} = -\sin\chi - \epsilon r \sin\theta_W \cos\chi, ~~U_{33} = r \cos\chi, 
\end{align}
where $\epsilon$ is given by Eq.~\eqref{eq:loop-induced-mix} and $\chi$ is the mixing angle of $Z$ and $Z'$ defined by
\begin{align}
\tan 2\chi =  - \frac{2 \epsilon r \sin\theta_W m_{Z}^2}{m_Z^2 - m_{Z'}^2}.
\label{eq:tan2chi}
\end{align}
Here we assume that the mixing angle $\chi$ is much smaller than unity, and hence the masses and eigenstates 
of the gauge bosons are approximately  the same as those without the kinetic mixing.

The decay width of $Z' \to f\bar{f}$ is given by
\begin{align}
\Gamma(Z' \to f \bar{f}) = \frac{m_{Z'}}{48 \pi} \beta(y_f) 
  \left[ {v'_f}^2 + {a'_f}^2 + 2({v'_f}^2 - 3 {a'_f}^2) \frac{m_f^2}{m_{Z'}^2}\right], \label{eq:gamma-Zp-ff}
\end{align}
where $y_f = m_f^2/\mZp^2$. In the limit of $\epsilon \to 0$, $v_f'$ is given by $2g' X_f$ and $a_f'$ vanishes.
In such a situation, the decay widths Eq.~\eqref{eq:gamma-Zp-ff} is proportional to $g'^2 \mZp$.

%%%%%%%%%%%%%%%%%%%%%%%%%%%%%%%%%%%%%%%%%%%%%%%%
\begin{figure}[t]
\begin{center}
	\includegraphics[width=0.6\textwidth]{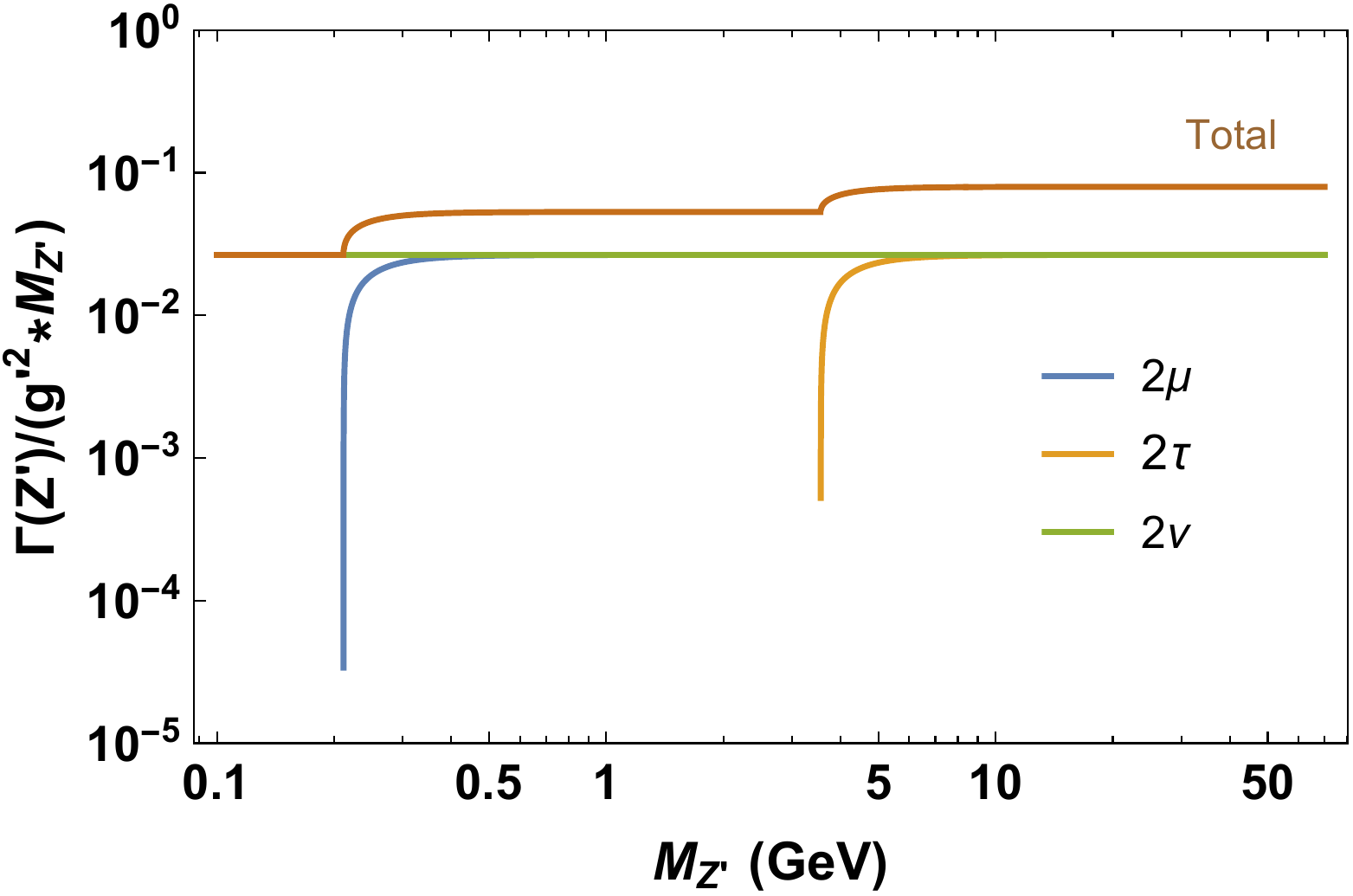} 
\end{center}
\caption{Decay width of $Z'$ normalized by ${g'}^2 \mZp$, $\Gamma(Z')/({g'}^2 m_{Z'})$ for various $g'$ 
as a function of $\mZp$. The blue, yellow and green curves represent the decay into muons, taus and neutrinos, 
respectively. The brown curve corresponds to the sum of the decay widths. In the figure, the gauge coupling constant 
is fixed to $10^{-3}$.}
\label{fig:Gamma-Zp-dep}
\end{figure}
%%%%%%%%%%%%%%%%%%%%%%%%%%%%%%%%%%%%%%%%%%%%%%%%%%%
Figure \ref{fig:Gamma-Zp-dep} shows the $\mZp$ dependence of the decay widths normalized by $g'^2 \mZp$.  
In the figure, $g'$ is fixed to be $10^{-3}$ and the kinetic mixing is set to be $10^{-3} g'$ which correspond to the loop induced kinetic mixing for $\mZp = 20$ GeV. 
The blue, yellow and green curves represent the decays into $\mu\mu,~\tau\tau$ and $\nu_\mu \nu_\mu + \nu_\tau \nu_\tau$. 
The brown one denotes the total widths.
From the figure, one can see that $Z'$ mainly decays into 
$\mu\mu,~\tau\tau$ and $\nu_\mu \nu_\mu+\nu_\tau \nu_\tau$, and the normalized decay widths are almost 
constant above the threshold of muon and tau. This is due to the fact that $v_f' \sim 2 g' X_f \gg a_f'$ 
in Eq.~\eqref{eq:coup-Zprime-ff}. Compared with $U_{33}$, the mixing matrix $U_{13}$ is much suppressed 
with $\epsilon \sim 10^{-3} g'$ and $U_{23}$ is also suppressed because of the cancellation in
$\sin\chi+\epsilon r \sin\theta_W \cos\chi$. Thus, the partial decay widths into $\mu\mu,~\tau\tau$ 
and $\nu_{\mu,\tau} \nu_{\mu,\tau}$ are almost proportional to $g'^2\mZp$.
The normalized decays widths into quarks and electrons are below about $10^{-7}$, and hence the 
branching ratio of these decays are below $10^{-6}$. 
Such a small portion of the decays will not be seen due to large background of the SM processes. Therefore, we only consider the decays into $\mu\mu$ as the 
signal in the following sections. Before closing this subsection, we should comment on the gauge 
mixing angle Eq.~\eqref{eq:tan2chi}. 
The gauge mixing angle increases as $\mZp$ get close to $m_Z$ and becomes maximal when $\mZp \simeq m_Z$. 
In such situation, $U_{23}$ is not suppressed and the $Z'$ interactions with quarks and electron is significant.  
Then, the decays into quarks and electrons can dominate $Z'$ decays. 
However, the mixing angle is kept below $\chi/g' \simeq  1.2 \times 10^{-3}$ for $\mZp \lsim 60$ GeV and $\epsilon/g' = 10^{-3}$. 
Therefore, the increase of the gauge mixing angle can be safely ignored in the decays of $Z'$.

%%%%%%%%%%%%%%%%%%%%%%%%%%%%%%%%%%%%%%%%%%%%%%%
\subsection{Decays of $h$} \label{subsec:decay-h}
%%%%%%%%%%%%%%%%%%%%%%%%%%%%%%%%%%%%%%%%%%%%%%%
The SM-like Higgs can decay into $Z'Z'$ and $\phi\phi$ when these are kinematically allowed.

The relevant Lagrangian of those new decay modes is given by
\begin{align}
\mathcal{L}_{h~\mathrm{int}} = \frac{\mZp}{v_\varphi} \sin\alpha  h Z'_\mu Z^{'\mu} -\lambda_{h 2\phi} h \phi^2,
\end{align}
where 
\begin{align}
\lambda_{h 2\phi} =- \frac{\sin 2\alpha (m_h^2 + 2 m_\phi^2) ( \cos \alpha v + \sin \alpha v_\varphi)}{4 v v_\varphi}.
\end{align}

The decay widths are given by
\begin{subequations}
\begin{align}
\Gamma(h \to Z' Z') &= \frac{g_{Z'Z'h}^2}{8 \pi m_h}  \beta(z_{Z'})
 \left[ 2 + \frac{m_h^4}{4 m_{Z'}^4} \left( 1 - \frac{2 m_{Z'}^2}{m_h^2} \right)^2 \right], \\
\Gamma(h \to \phi \phi) &= \frac{\lambda_{2h \phi}^2}{8 \pi m_h} \beta(z_\phi),
\end{align}
\end{subequations}
where $z_{Z'} = \mZp^2/m_h^2$.

The invisible decay branching ratio of $h$ is constrained at the LHC experiment. In this model, the invisible decays are 
\begin{subequations}
\begin{align}
h &\to Z' Z' \to 4 \nu, \\
h &\to ZZ' \to 4 \nu, \\
h &\to \phi \phi \to 4 Z' \to 8 \nu.
\end{align}
\end{subequations}
The branching ratio of these decays is given by
\begin{align}
\mathrm{Br}(h \to \mathrm{invisibles}) &= 
	\mathrm{Br}(h \to Z'Z') \big[ \mathrm{Br}(Z' \to \nu \nu) \big]^2
	+ \mathrm{Br}(h \to ZZ') \big[ \mathrm{Br}(Z \to \nu \nu) \mathrm{Br}(Z' \to \nu \nu) \big] \nonumber \\
&\quad + \mathrm{Br}(h \to \phi\phi) \big[ \mathrm{Br}(\phi \to Z' Z') \big]^2 \big[ \mathrm{Br}(Z' \to \nu \nu) \big]^4,
\end{align}
where its current bound is less than $0.25$ and $0.19$ by the CMS~\cite{Sirunyan:2018owy} and the ATLAS~\cite{Aaboud:2019rtt} experiments. To be conservative, we employ the bound from the CMS experiment 
in the following discussion.

The decays of $h$ into charged leptons are also constrained at the LHC. Its constraint is more stringent when 
the branching ratio of $Z' \to l\bar{l}$ is sizable.

The new decay processes into charged leptons in this model are 
\begin{subequations}
\begin{align}
h &\to Z' Z' \to l_1 \overline{l_1} + l_2 \overline{l_2}, \label{eq:decay-h-4l-1} \\
h &\to Z Z' \to  l_1 \overline{l_1} + l_2 \overline{l_2}, \label{eq:decay-h-4l-2} \\
h &\to \phi \phi \to 4 Z' \to l_1 \overline{l_1} + l_2 \overline{l_2} + l_3 \overline{l_3} + l_4 \overline{l_4}, \label{eq:decay-h-8l}
\end{align}
\end{subequations}
where $l= e,~\mu,~\tau$. 

In \cite{Aad:2015sva}, a light gauge boson search was performed and impose stringent constraint on $m_{Z'}$ and $g'$. 
The branching ratio of \eqref{eq:decay-h-4l-1} and \eqref{eq:decay-h-4l-1} is given by
\begin{align}
\mathrm{Br}(h \to 4l) &=
	\mathrm{Br}(h \to Z'Z') \big[2  \mathrm{Br}(Z' \to l_1 \overline{l_1})  \mathrm{Br}(Z' \to l_2 \overline{l_2}) \big] \nonumber \\
&\quad + \mathrm{Br}(h \to ZZ') \big[ \mathrm{Br}(Z \to l_1 \overline{l_1})  \mathrm{Br}(Z' \to l_2 \overline{l_2}) 
	 + \mathrm{Br}(Z' \to l_1 \overline{l_1})  \mathrm{Br}(Z \to l_2 \overline{l_2}) \big].
\label{eq:Br-h-to-4l}
\end{align}

%%%%%%%%%%%%%%%%%%%%%%%%%%%
\section{Constraints}
%%%%%%%%%%%%%%%%%%%%%%%%%%%
We explain the relevant constraints on the gauge coupling $g'$, the $Z'$ boson mass $\mZp$ and the scalar mixing $\alpha$, 
and show the allowed region for these parameters. In the following discussion, we assume that the extra scalar boson is 
heavier than the Higgs boson so that the decay of $h \to \phi \phi$ is kinematically forbidden.
The ATLAS collaboration searched for new light gauge bosons via the Higgs boson decays into $4$ 
leptons, $ h \to Z'Z' \to 2l 2l'$ and $h \to ZZ' \to 2l 2l'$, \cite{Aad:2015sva,Aaboud:2018fvk}. In our setup, the latter decay 
is much suppressed by the loop-induced kinetic mixing, and hence the bound does not restrict the parameters. 
The former decay, on the other hand, occurs through the $\lmlt$ gauge interaction and the scalar mixing, 
which constrains the parameters.  
The $1\sigma$ and $2\sigma$ upper limit (UL) on the decay branching ratio of $h \to Z'Z' \to 4l$ are shown 
in Table \ref{tab:upper-lim-higgs-into-4lep}. 
%%%%%%%%%%%%%%%%%%%%%%%%%%%%%%%%%%%%%%%%%
\begin{table}[tbh]
\begin{center}
\begin{tabular}[t]{|c|c|c||c|} \hline
                                   & \multicolumn{2}{|c||}{ Br$(h \to Z'Z' \to 4l)$} &  $g'$ \\ \hline
~~$m_{Z'}$ (GeV)~~ & ~~~~~$1\sigma$ UL ~~~~~ & ~~~~~ $2\sigma$ UL ~~~~~ & ~~~~~ $95$\% CL ~~~~~ \\ \hline \hline 
$20.0$ & $9.4 \times 10^{-5}$  & $1.1 \times 10^{-4}$  & $5.8 \times 10^{-3}$ \\ \hline
$30.0$ & $9.5 \times 10^{-5}$  & $1.4 \times 10^{-4}$  & $1.2 \times 10^{-2}$ \\ \hline
$40.0$ & $1.2 \times 10^{-4}$  & $1.6 \times 10^{-4}$  & $2.5 \times 10^{-2}$ \\ \hline
$50.0$ & $1.5 \times 10^{-4}$  & $2.2 \times 10^{-4}$  & $3.9 \times 10^{-2}$ \\ \hline
$60.0$ & $1.9 \times 10^{-4}$  & $2.8 \times 10^{-4}$  & $9.6 \times 10^{-2}$ \\ \hline
\end{tabular}
\end{center}
\caption{
Upper limit on the branching ratios of the Higgs boson decays into $Z'Z'$ followed by $Z' \to l\bar{l}$ and the 
gauge coupling constant $g'$.  
Numbers of the branching ratio and the gauge coupling are taken from Figure 10.(a) of \cite{Aaboud:2018fvk} and Figure 8 of \cite{Sirunyan:2018nnz}, respectively. 
}
\label{tab:upper-lim-higgs-into-4lep}
\end{table}
%%%%%%%%%%%%%%%%%%%%%%%%%%%%%%%%%%%%%%%%%%%%%%%%%%%%%

The CMS collaboration also searched the $Z'$ boson using $77.3$ fb$^{-1}$ data recorded in $2016$ 
and $2017$ \cite{Sirunyan:2018nnz}. 
The search was performed for the bremsstrahlungs of $Z'$ from $\mu$ or $\Am$ produced in $pp$ collisions.
The results put the constraint on $g'$ in the mass range $5 \leq \mZp \leq 70$ GeV.
The upper bound on $g'$ is shown in Table \ref{tab:upper-lim-higgs-into-4lep}.

\begin{figure}[t]
\begin{center}
\begin{tabular}{c}
	\includegraphics[height=80mm]{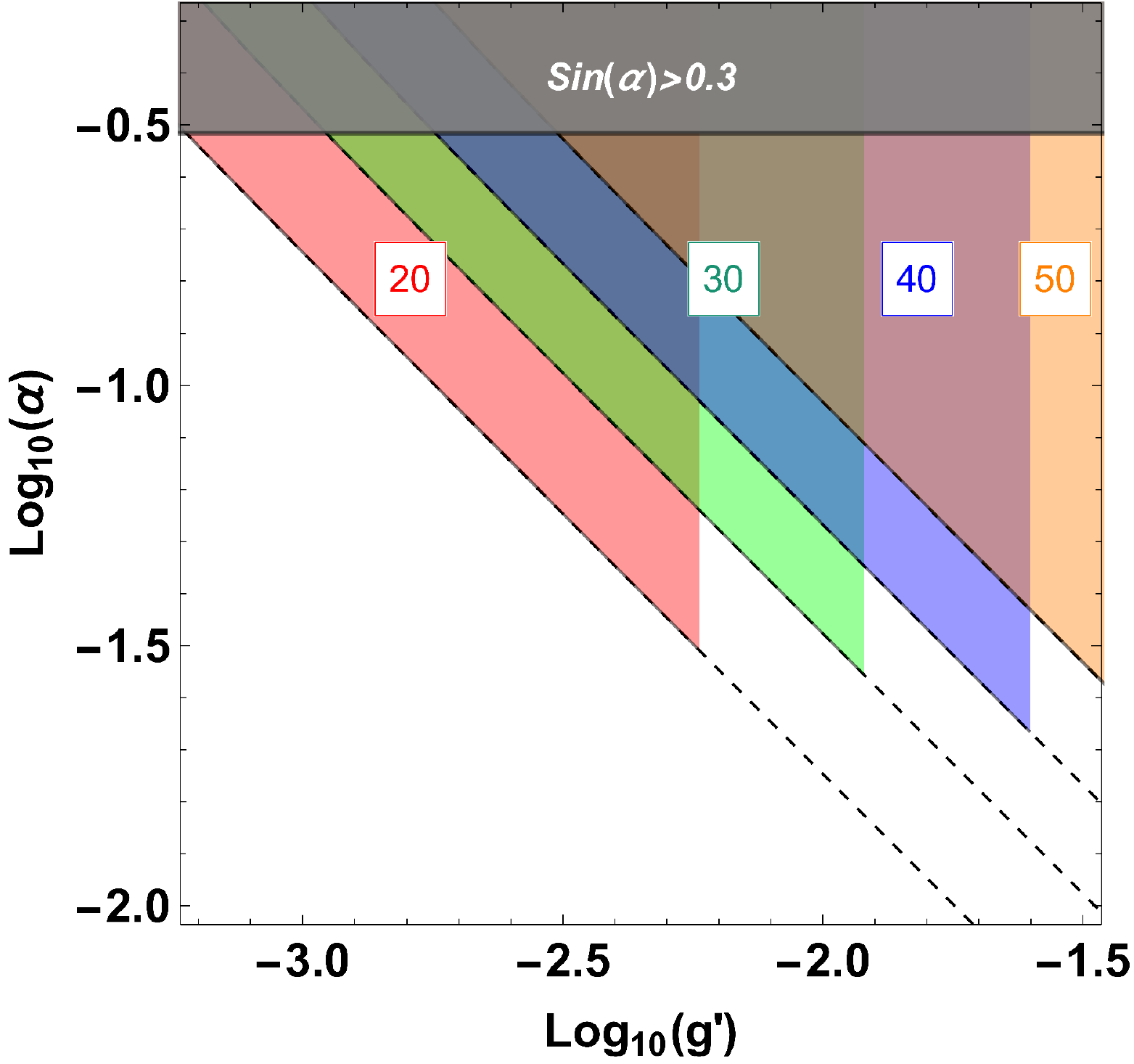} \end{tabular}
\end{center}
\caption{ Allowed region of the parameters in $\log(g')$-$\log(\alpha)$ plane. Red, green, blue and orange colored region are 
the exclusion region for $\mZp=20,~30,~40$ and $50$ GeV from $h \to 4l$ search at $2\sigma$. Dashed lines represent  
the contour of the upper bound of Br$(h \to 4l)$ given in Table \ref{tab:upper-lim-higgs-into-4lep}. Gray shaded region is 
excluded by analysis of data regarding the SM Higgs signals from the LHC experiments.
}
\label{fig:allowed-region-gp-alpha}
\end{figure}

From the experimental bounds explained above, we calculate the branching ratio of $h \to 4l$, Eq.~\eqref{eq:Br-h-to-4l}, 
and showed the allowed region for $g'$ and $\alpha$ in Figure \ref{fig:allowed-region-gp-alpha}.
%%%%%%%%%%%%%%%%%%%%%
%%%%%%%%%%%%%%%%%%%%%
Red, green, blue and and orange filled region are the exclusion region for $\mZp=20,~30,~40$ and $50$ GeV, from the search for $h \to 4l$ at $2\sigma$, respectively. Dashed lines are the contour of the upper limit of the branching ratio. Gray region is excluded 
by analysis of data regarding the SM Higgs signals from the LHC experiments~\cite{Cheung:2015dta,Chpoi:2013wga}.

%%%%%%%%%%%%%%%%%%%%%%%%%%%
\section{$Z'$ and $\phi$ search at LHC \label{sec:search}}
%%%%%%%%%%%%%%%%%%%%%%%%%%%

In this section we carry out numerical simulation for $\phi$ production followed by $\phi \to Z'Z'$ and $Z' \to \mu^+ \mu^-$ decays at the LHC.
The background (BG) process, $pp \to \mu^+ \mu^- \mu^+ \mu^-$, is also considered in the SM.
Then significance of the signal is discussed applying relevant kinematical cuts where experimental constraints in previous section are also 
taken into account.

%%%%%%%%%%%%%%%%%%%%%%%%%%%
\subsection{Production Cross Section}
%%%%%%%%%%%%%%%%%%%%%%%%%%%
%%%%%%%%%%%%%%%%%%%
\begin{figure}[t]
\begin{center}
\includegraphics[width=70mm]{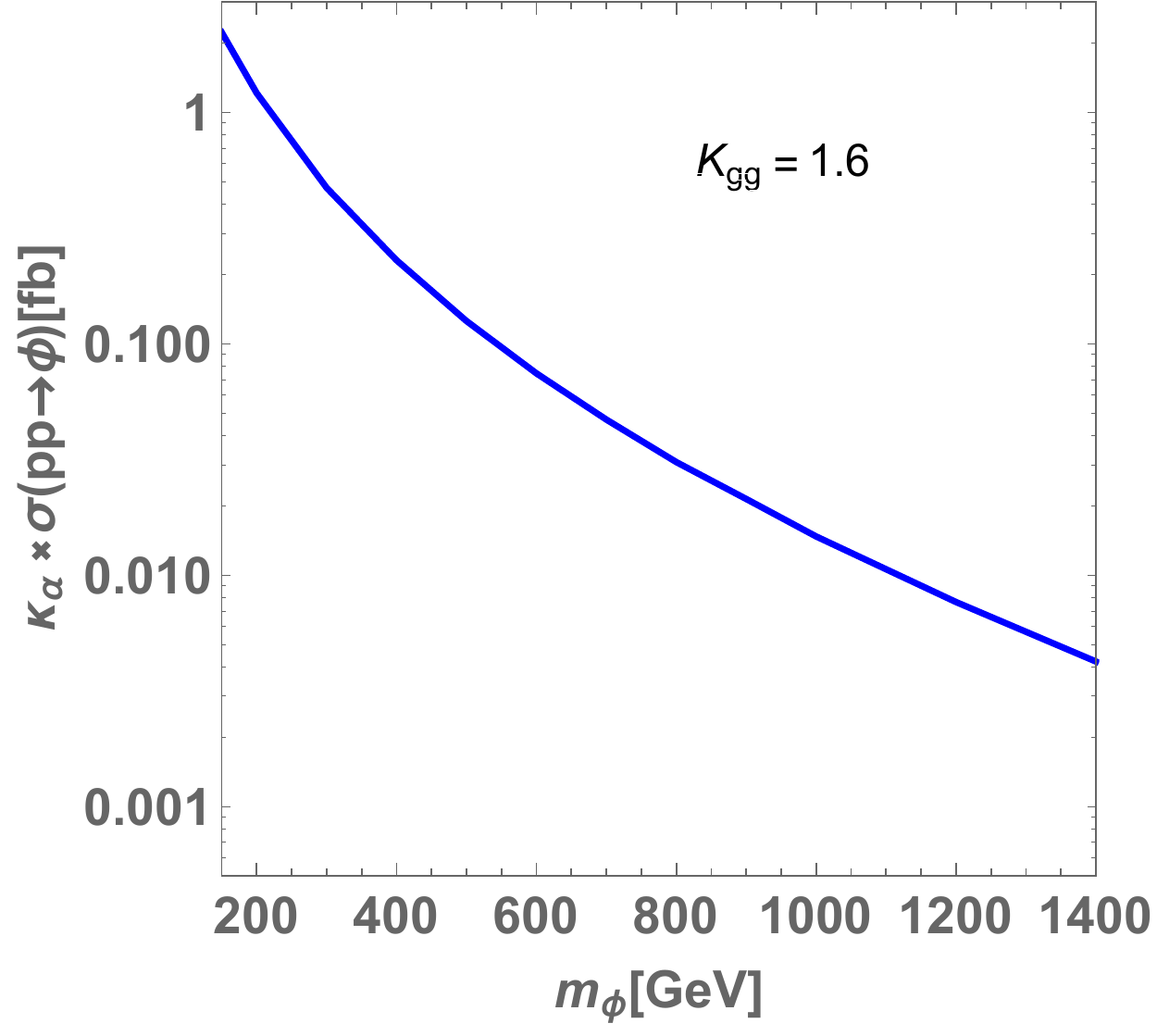}
\caption{The cross section for $pp \to \phi$ as a function of $m_\phi$ which is multiplied by scaling factor $\kappa_\alpha = (\sin \alpha/0.01)^{-2}$, and $\sqrt{s} = 14$ TeV is applied.} 
  \label{fig:CX}
\end{center}\end{figure}
%%%%%%%%%%%%%%%%%%

Here we discuss the dominant $\phi$ production process at the LHC. 
This new scalar boson can be produced by gluon fusion process $gg \to \phi$ through mixing with the SM Higgs boson.
We obtain the relevant effective interaction for the gluon fusion as~\cite{Gunion:1989we}
\begin{equation}
{\mathcal L}_{\phi gg} = \frac{\alpha_s}{16 \pi} \frac{\sin \alpha}{v} A_{1/2}(\tau_t) \phi G^a_{\mu \nu} G^{a \mu \nu}, 
 \end{equation}
 where $A_{1/2}(\tau_t) = -\frac{1}{4} [\ln[(1+\sqrt{\tau_t})/(1-\sqrt{\tau_t})] - i \pi ]^2$ with $\tau_t = 4 m_{t}^2/m_\phi^2$ and  $G^a_{\mu \nu}$ is the field strength for gluon.
 This effective interaction is dominantly induced from $\bar t t \phi$ coupling via the mixing effect where we take into account only top Yukawa coupling omitting the other subdominant contributions for simplicity.
In Fig.~\ref{fig:CX}, we show the production cross section as a function of $m_\phi$ estimated by use of {\tt MADGRAPH5}~\cite{Alwall:2014hca} implementing the effective interaction wit FeynRules 2.0 \cite{Alloul:2013bka}, which is multiplied by scaling factor $\kappa_\alpha \equiv (\sin \alpha/0.01)^{-2}$ as the cross section is proportional to $\sin^2 \alpha$.
In addition we included K-factor $K_{gg} = 1.6$ for gluon fusion process which represent NLO correction effect~\cite{Djouadi:2005gi}.

%%%%%%%%%%%%%%%%%%%%%%%%%%%
\subsection{Kinematical Cuts}
%%%%%%%%%%%%%%%%%%%%%%%%%%%

 %%%%%%%%%%%%%%%%%%%%%%%%%%%%%
\begin{figure}[t]
\centering
\includegraphics[width=80mm]{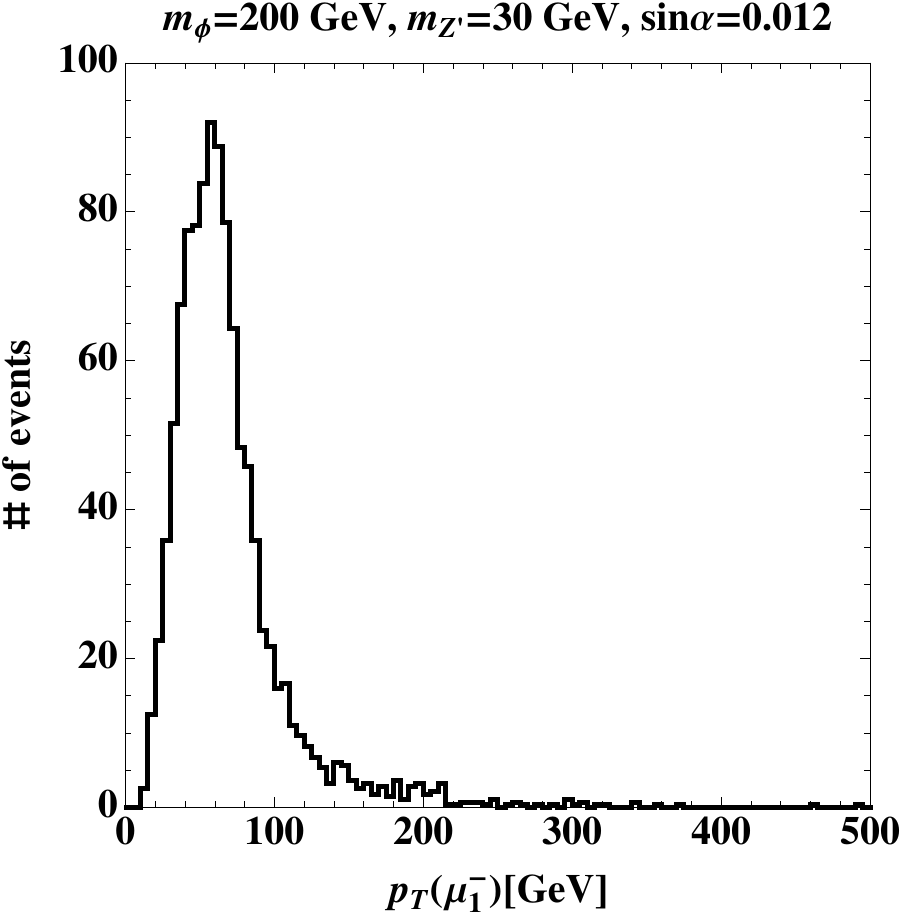} 
\includegraphics[width=80mm]{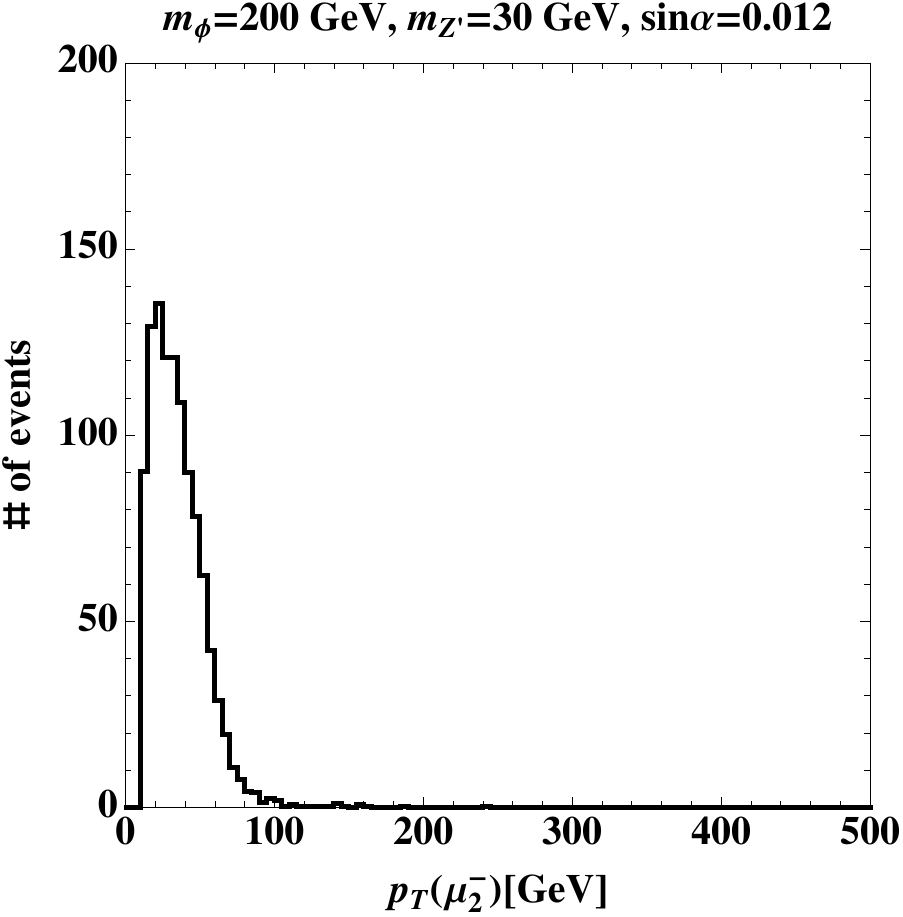} \\
\includegraphics[width=80mm]{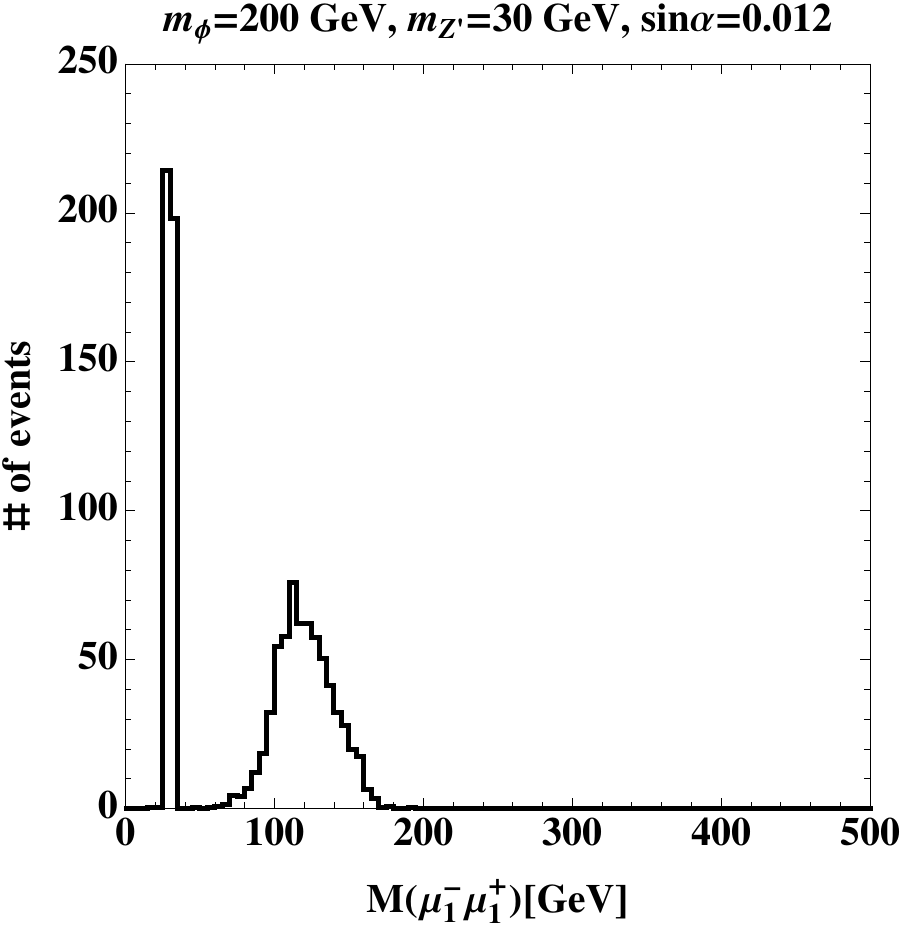} 
\includegraphics[width=80mm]{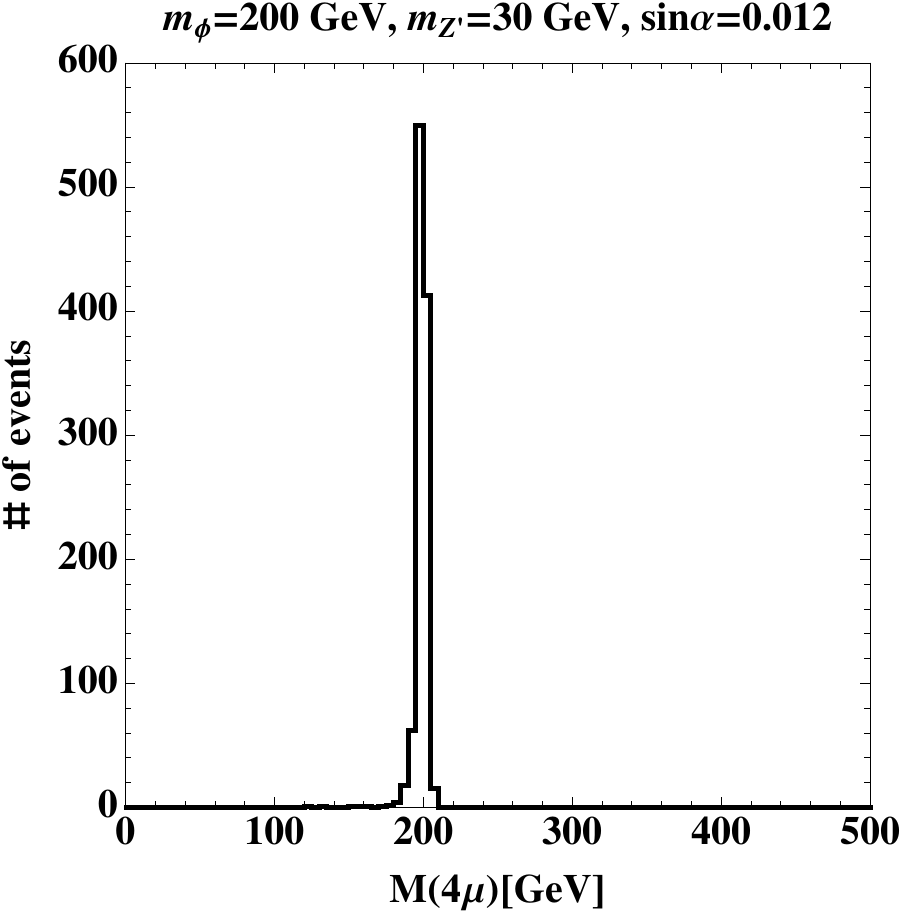} 
\caption{Kinetic distributions of final state muons for signal.}
\label{fig:distribution-sig}
\end{figure}
%%%%%%%%%%%%%%%%%%%

 %%%%%%%%%%%%%%%%%%%%%%%%%%%%%
\begin{figure}[t]
\centering
\includegraphics[width=80mm]{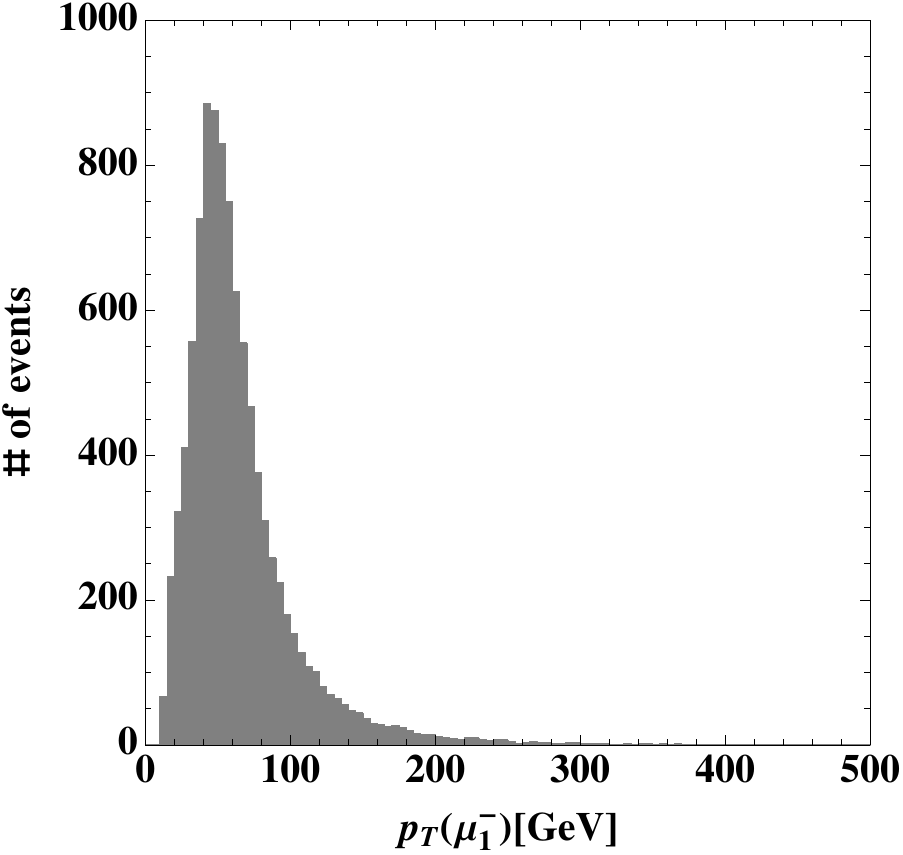} 
\includegraphics[width=80mm]{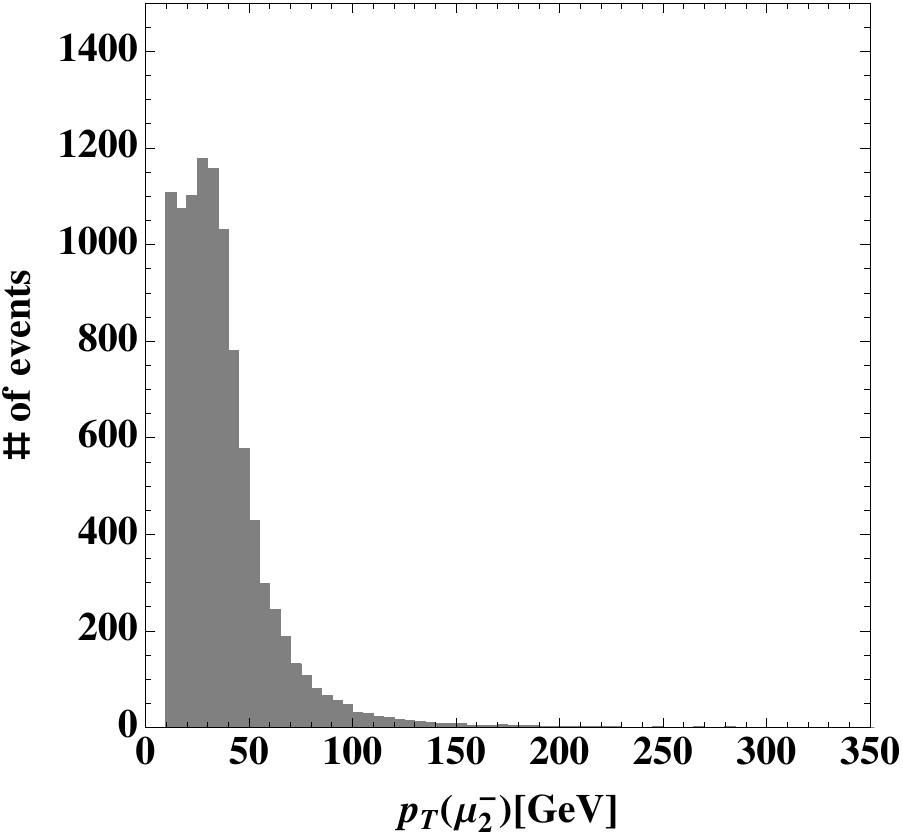} \\
\includegraphics[width=80mm]{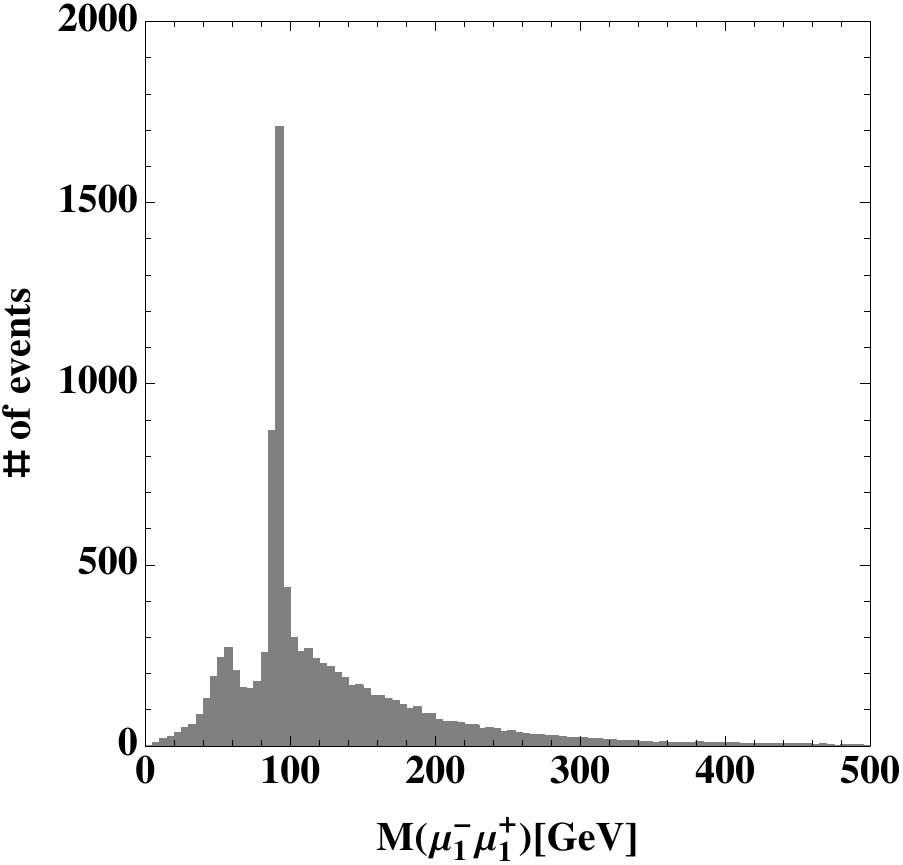} 
\includegraphics[width=80mm]{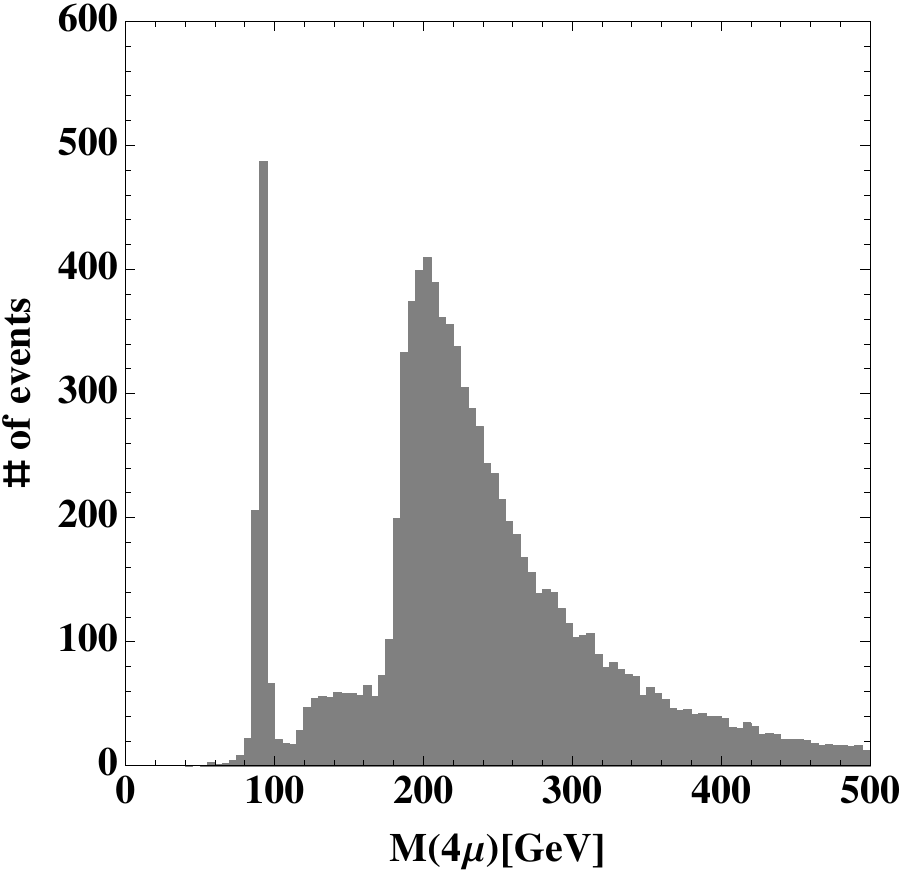} 
\caption{Kinetic distributions of final state muons for BG.}
\label{fig:distribution-BG}
\end{figure}
%%%%%%%%%%%%%%%%%%%

We carry out numerical simulation for our signal and BG processes where the events are generated using {\tt MADGRAPH/MADEVENT\,5}~\cite{Alwall:2014hca} implementing the necessary Feynman rules and relevant parameters of the model via FeynRules 2.0 \cite{Alloul:2013bka}, 
the {\tt PYTHIA\,8}~\cite{Sjostrand:2014zea}  is applied to deal with hadronization effects,  the  initial-state radiation (ISR) and final-state radiation (FSR) effects and the decays of SM particles, and {\tt Delphes}~\cite{deFavereau:2013fsa} is used for detector level simulation.
In generating signal and BG events basic cuts are implemented in {\tt MADGRAPH/MADEVENT\,5} as 
\begin{equation}
\label{eq:BasicCutsL}
p_T(\ell^\pm) > 7 \ {\rm GeV}, \quad \eta(\ell^\pm) < 2.5,
\end{equation} 
where $p_T$ denotes transverse momentum and $\eta = 1/2 \ln (\tan \theta/2)$ is the pseudo-rapidity given by $\theta$ being the scattering angle in the laboratory frame.
We then chose events which has two muon anti-muon pair in final states. 

To impose additional cuts we produce kinematical distributions for signal and BG.
In Figs.~\ref{fig:distribution-sig} and \ref{fig:distribution-BG} we show several distributions for signal and BG 
where we fix $m_\phi = 200$ GeV, $m_{Z'} = 30$ GeV and $\sin \alpha =0.012$ as reference values for signal events.
In addition we chose integrated luminosity as 3000 fb$^{-1}$ in estimating number of events.
The upper-left and -right plots in the figures show distributions for transverse momentum of $\mu^-$ where $\mu_1^-$ and $\mu^-_2$ are distinguished by $p_T(\mu^-_1) > p_T(\mu^-_2)$ for each event;
the distributions for $\mu^+$ are the same as $\mu^-$.
We find that signal and BG provide similar distribution for transverse momentum of muon.
Thus we do not impose further cuts for $p_T(\mu)$.
The lower-left plots in the figures show distributions for invariant mass of $\mu_1^- \mu_1^+$.
For signal we find clear peak corresponding to $Z'$ mass where another bump comes from combinations of muon from different $Z'$ decays. 
On the other hand peak at $Z$ mass is found for BG.
The lower-right plots in the figures show distributions for invariant mass of four muons $M_{\mu^+\mu^+\mu^-\mu^-}$.
For signal the distribution is clearly concentrated at the mass of $\phi$ while the distribution for BG shows peak at $Z$ mass and continuous region.
To reduce the BG events, we thus eliminate events which has  $\mu^+ \mu^-$ and  $\mu^+\mu^+\mu^-\mu^-$ invariant masses within the range of 
\begin{align}
80 \ {\rm GeV} < M_{\mu^+ \mu^-} < 100 \ {\rm GeV}, \\
80 \ {\rm GeV} < M_{\mu^+\mu^+\mu^-\mu^-} < 100 \ {\rm GeV}.
\end{align}
We estimate numbers of signal and BG events after applying these cuts.

%%%%%%%%%%%%%%%%%%%%%%%%%%%
\subsection{Significance}
%%%%%%%%%%%%%%%%%%%%%%%%%%%

 %%%%%%%%%%%%%%%%%%%%%%%%%%%%%
\begin{figure}[t]
\centering
\includegraphics[width=80mm]{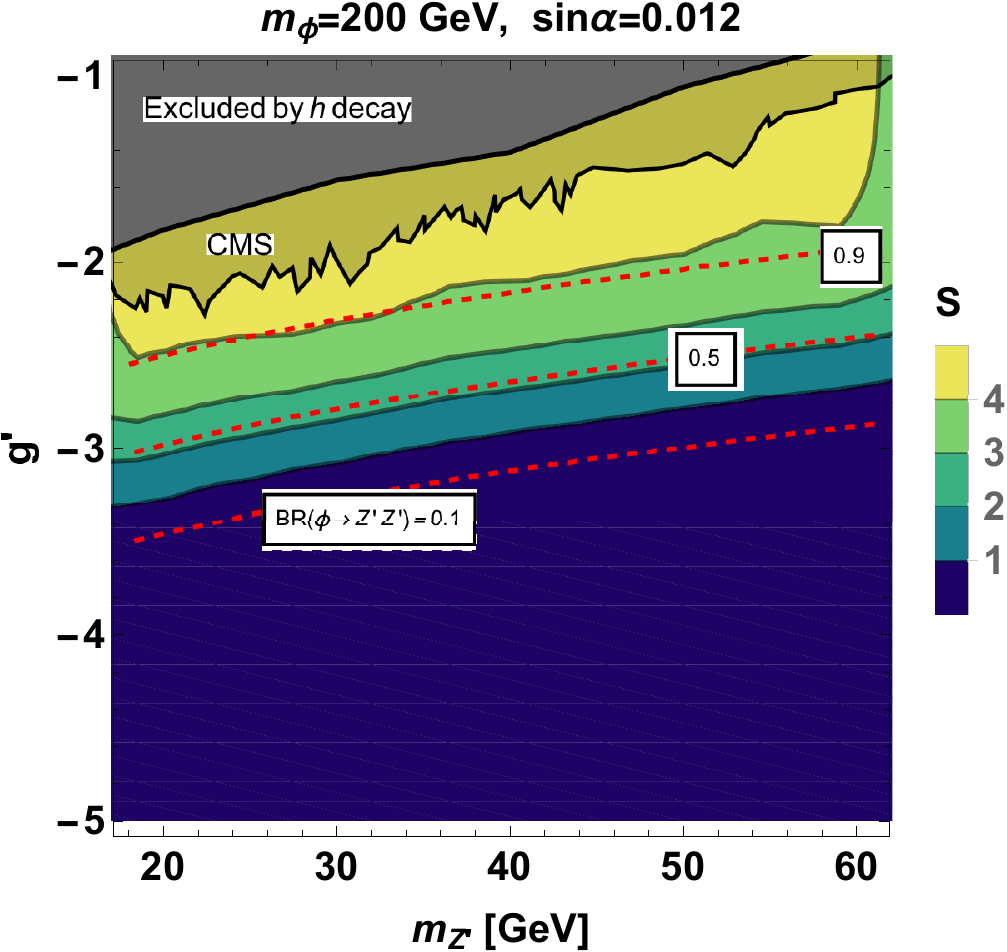} 
\includegraphics[width=80mm]{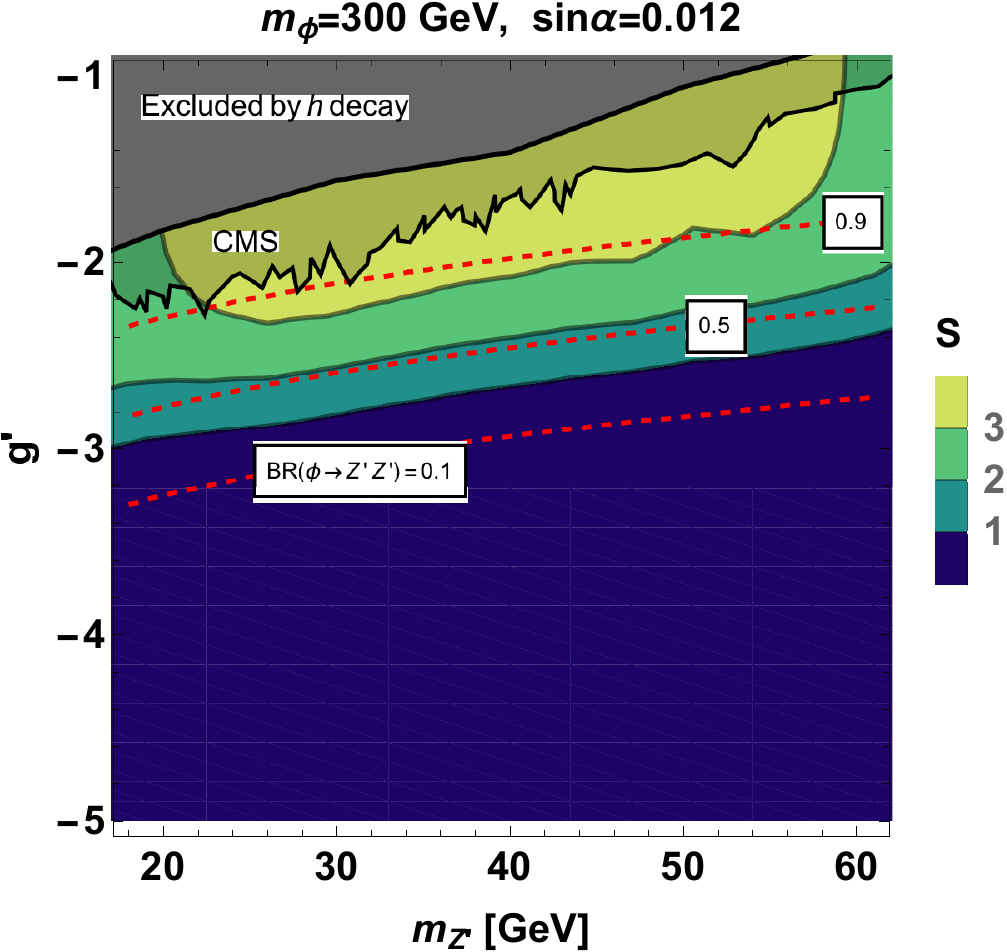} \\ \vspace{5mm}
\includegraphics[width=80mm]{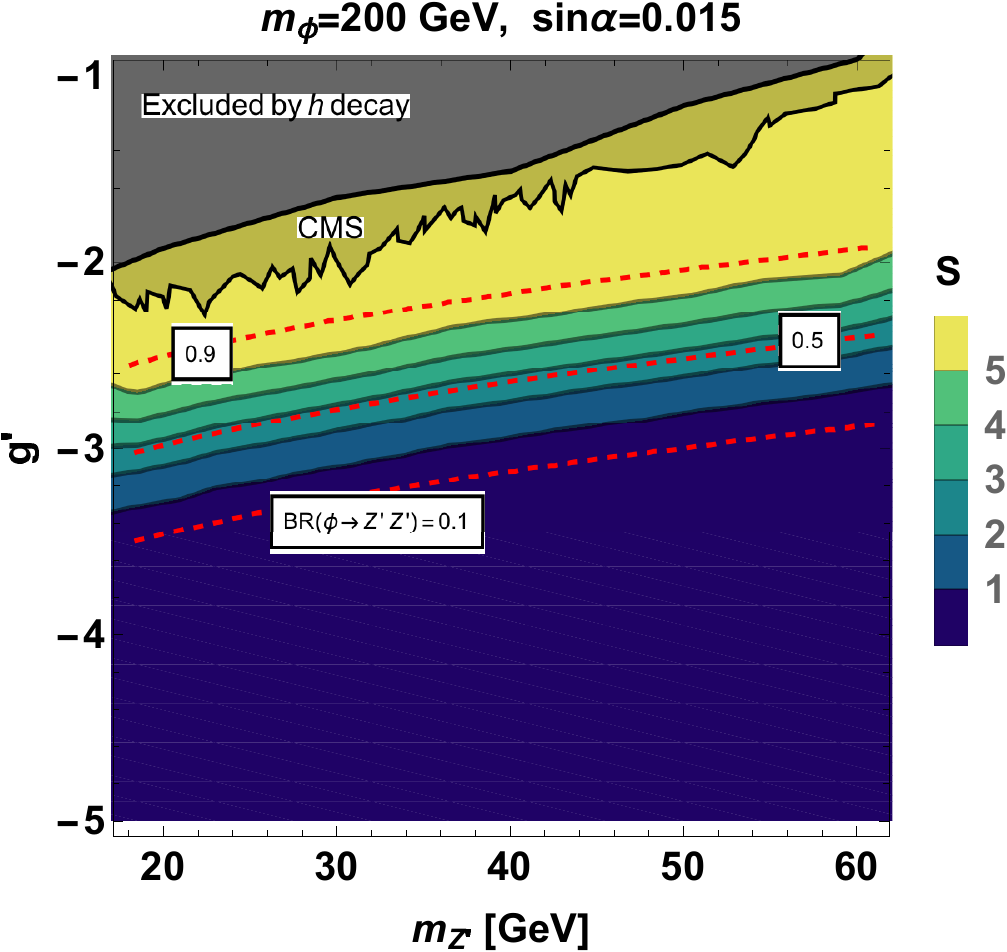} 
\includegraphics[width=80mm]{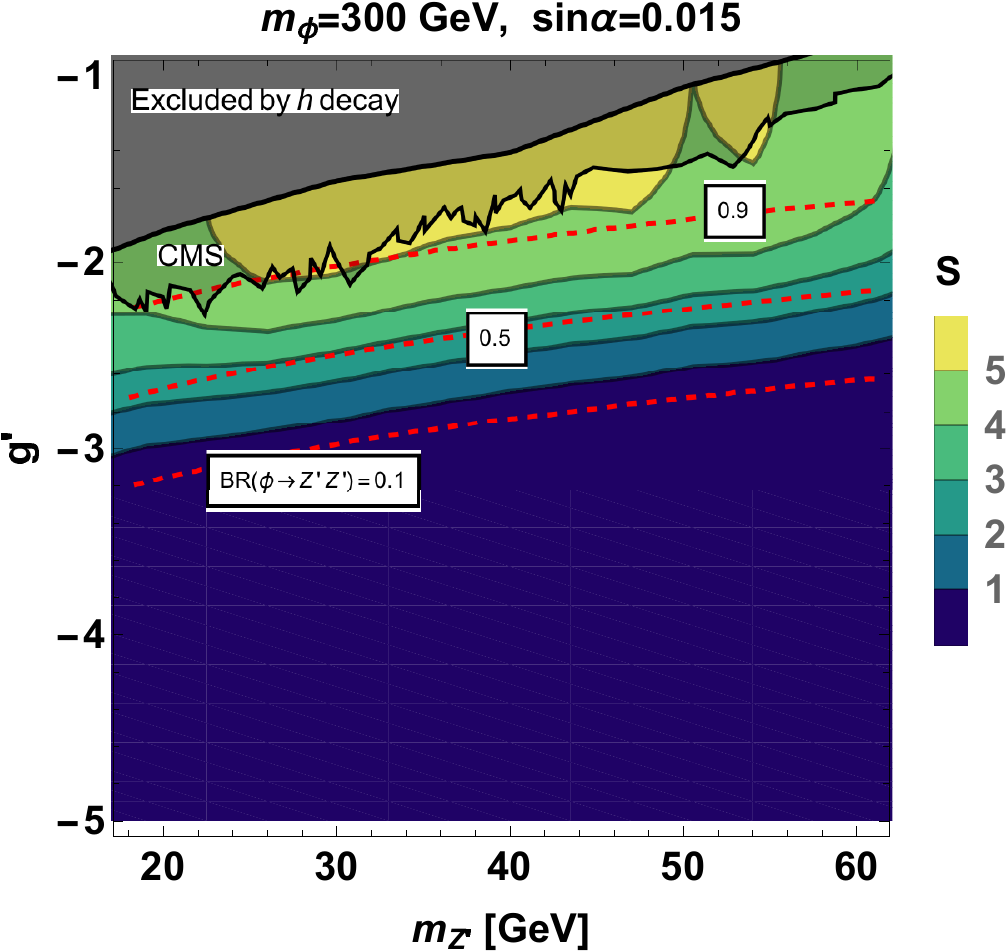} 
\caption{Significance on $m_{Z'}$--$g'$ plane where red dashed lines are for $BR(\phi \to Z' Z')$ and its values are indicated on the lines.}
\label{fig:significance}
\end{figure}
%%%%%%%%%%%%%%%%%%%
 %%%%%%%%%%%%%%%%%%%%%%%%%%%%%
\begin{figure}[t]
\centering
\includegraphics[width=80mm]{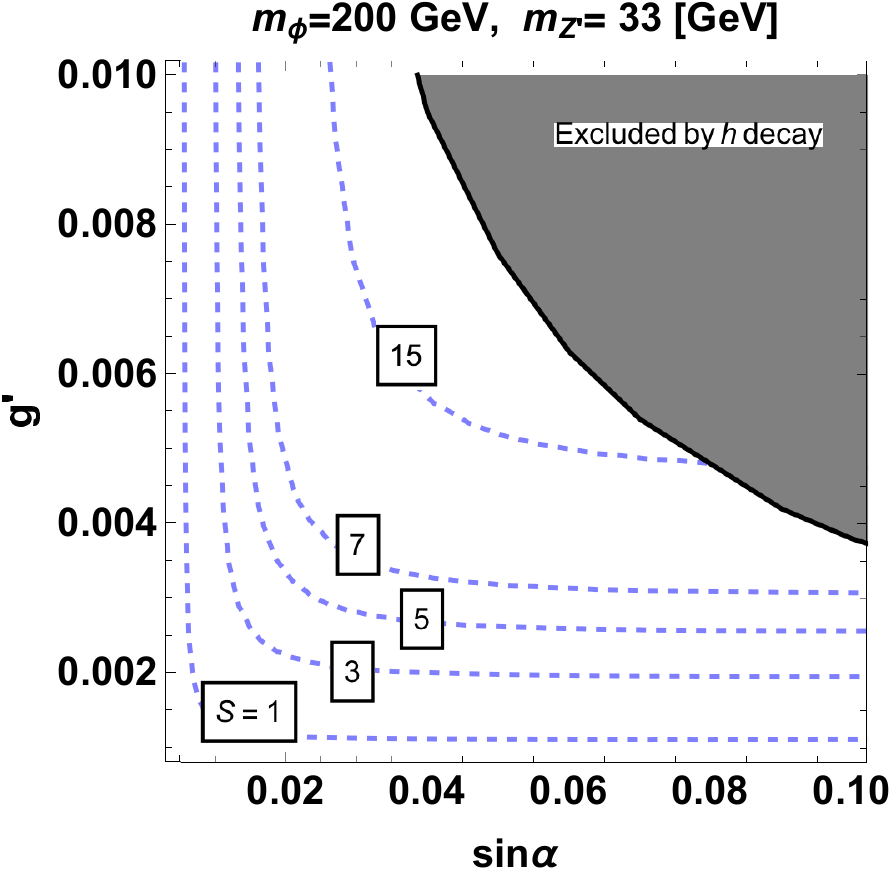} 
\includegraphics[width=80mm]{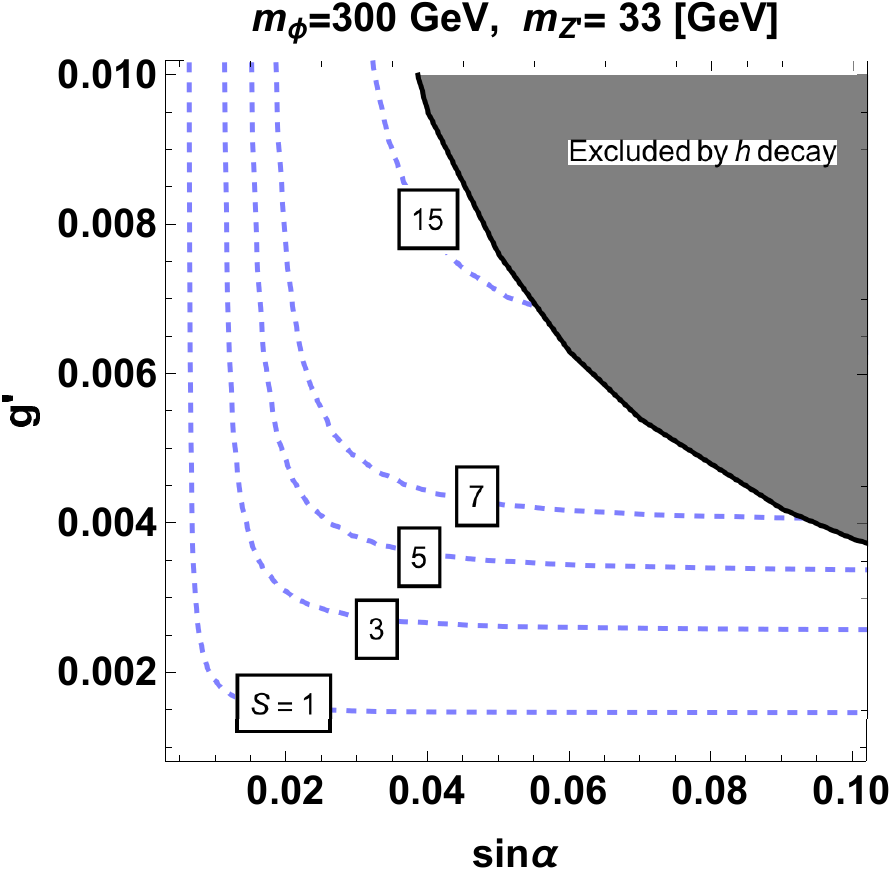} 
\caption{Significance on $\sin \alpha$--$g'$ plane.}
\label{fig:significance2}
\end{figure}
%%%%%%%%%%%%%%%%%%%

We estimate discovery significance of the signal after imposing kinematical cuts discussed in previous subsection.
The significance is given by
\begin{equation} 
\label{eq:sig}
S = \frac{N_S}{\sqrt{N_{BG}}},
\end{equation}
where $N_S$ and $N_{BG}$ are respectively the number of events for signal and total BG.
In estimating number of events we assume integrated luminosity of 3000 fb$^{-1}$ as in the previous subsection.
In Fig.~\ref{fig:significance} we show contours of discovery significance on $\{m_{Z'}, g' \}$ plane 
where we fix $\{m_\phi, \sin \alpha \}=\{200 \ {\rm GeV}, 0.012\}$, $\{300 \  {\rm GeV}, 0.012\}$, $\{200 \ {\rm GeV}, 0.015\}$ and $\{300 \ {\rm GeV}, 0.015\}$  as reference values for upper-left, upper-right, lower-left and lower-right plots.
In addition value of $BR(\phi \to Z'Z')$ is shown by red dashed curve, gray region is excluded by the constraint from $h \to Z' Z' \to \ell^+ \ell^- \ell^+ \ell^-$ decay, 
and light gray region is excluded by CMS data for $pp \to \mu^+ \mu^- Z'(\to \mu^+ \mu^-)$ signal search.
We find that $S \gtrsim 3$ can be achieved on parameter space with $g' \gtrsim 10^{-3(-2)}$ for $m_\phi = 200(300)$ GeV case when $\sin \alpha =0.012$.
For $\sin \alpha = 0.015$, we can obtain $S > 5$ with $g' \gtrsim 10^{-3} - 10^{-2}$ as the $\phi$ production cross section becomes large.
Moreover we show contours of the significance on $\{\sin \alpha, g' \}$ plane fixing $m_{Z'} = 33$ GeV and $m_{\phi} = 200(300)$ GeV in left(right) plot of Fig.~\ref{fig:significance2}.
To obtain sizable significance $\sin \alpha$ should be larger than $\mathcal{O}(0.01)$ to achieve sufficiently large $\phi$ production cross section. 
Also we cannot obtain sizable significance for parameter region with too small $g'$ since BR of $\phi \to Z'Z'$ becomes tiny in such region.
For very small $g'$ region, it will be more promising to search for signal that $\phi$ decays into SM particles, but analysis of such signals is beyond the scope of this paper.

%%%%%%%%%%%%%%%%%%%%%%%%%%%
\section{Conclusion}
%%%%%%%%%%%%%%%%%%%%%%%%%%%
We have considered a minimal gauged $\lmlt$ model where the $\lmlt$ gauge symmetry is spontaneously broken 
by a SM singlet scalar field, and studied the possibility of discovering the new gauge boson $Z'$ 
and scalar boson $\phi$ at the LHC experiments. 
We considered the case in which $\phi$ is heavier than $Z'$ so that it dominantly decays into $Z'Z'$. 
The produced $Z'$ decays into a pair of muons, taus and their corresponding neutrinos. 
Then, the signal significance of such $\phi$ and $Z'$ decays against the SM backgrounds is analyzed focusing on 
the four muon final states.

We firstly showed the branching ratio of $\phi \to Z'Z'$ decay becomes larger as the scalar mixing is smaller. 
The scalar boson dominantly decays into $Z'Z'$ for $\alpha \sim \mathcal{O}(10^{-2})$ for $m_\phi = 20$ and $50$ GeV 
as reference parameters. The decay width of $Z'$ is also showed that those into muons, taus and neutrinos are the 
almost the same. Thus, the branching ratio of $Z'$ into muons is $1/4$ in our setup.
The gauge coupling constant and the scalar mixing have been constrained by the searches for the four lepton decay 
and the invisible decays of the Higgs boson. Based on these constraints, the allowed region of $g'$ and $\alpha$ 
was derived for the analyses of the signal significance at LHC.

Then, the production cross section of $\phi$ through gluon fusion at LHC was calculated and the kinematical 
distributions of muons for the signal and backgrounds were analyzed. 
We found that the mass of $Z'$ and $\phi$ can be clearly 
reconstructed as a peak in the distributions of the invariant mass of $\mu^+ \mu^-$ and $\mu^+ \mu^+ \mu^- \mu^-$, respectively, 
for the signal events. 
On the other hand, for the background, the mass of $Z$ boson can be reconstructed in the same distributions. 
By applying cuts vetoing $80$ GeV  $< M_{\mu^+ \mu^-},~M_{\mu^+ \mu^ +\mu^- \mu^-} < 100$ GeV, the background 
events can be reduced. We showed the signal significance $S$ can reach to $3$ for $m_\phi = 200~(300)$ GeV and
 $g' \geq 10^{-3~(-2)}$, respectively, when the scalar mixing is $0.012$. For $\alpha = 0.015$,  $S > 5$ can be achieved.
 We also showed the signal significance in $\sin\alpha$-$g'$ plane. The region for $\sin\alpha \geq 10^{-2}$ 
 and $g' \geq 10^{-3}$ can be explored with $S > 3$.

\section*{Acknowledgments}
%\begin{acknowledgements}
This work is supported by JSPS KAKENHI Grant No.~JP18K03651,~JP18H01210 and 
MEXT KAKENHI Grant No.~JP18H05543 (T.~S.).
%\end{acknowledgements}

%%%%%%%%%%%%%%%%%%
%%% references %%%
%%%%%%%%%%%%%%%%%%
\bibliographystyle{apsrev}
\bibliography{biblio}

\end{document}